\documentclass[journal=apchd5,manuscript=article]{achemso}

\usepackage[version=3]{mhchem} 
\usepackage{siunitx} 
\usepackage{amsmath}

\author{Ilia Lykov}
\author{Sergejs Boroviks}
\author{Olivier J. F. Martin}
\affiliation{Nanophotonics and Metrology Laboratory,
Swiss Federal Institute of Technology Lausanne (EPFL),
1015 Lausanne, Switzerland}
\email{olivier.martin@epfl.ch}

\title{Simulations of Fluorescence and Energy Transfer in Periodic Systems: Pitfalls and Considerations for Bound States in the Continuum}


\begin{document}

\begin{abstract}
Bound states in the continuum (BICs) combine high quality factors with strong electromagnetic-field localization and are therefore widely explored for enhancing light--matter interactions. However, large field amplitudes do not directly translate into useful fluorescence enhancement or energy transfer enhancement between two donor--acceptor fluorophores. Here, we investigate this question for a hybrid plasmonic--photonic resonant waveguide grating supporting a symmetry-protected BIC. We emphasize some of the pitfalls associated with the numerical simulations, especially the utilization of periodic boundary conditions for fluorescence. Strong BIC-related fields can enhance acceptor absorption while useful fluorescence enhancement remains moderate. Extensive simulations and the comparison of different figures of merit, indicate that, while a BIC can suppress radiative leakage, it can also redirect a large fraction of the energy into Ohmic losses. As a result, the useful fluorescence enhancement of donors remains moderate, whereas absorption by acceptors can be strongly enhanced in F\"orster resonance energy transfer (FRET) experiments. It is essential to separate near-field FRET from long-range energy transfer and reabsorption mechanisms, both of which are mediated by the delocalized BIC mode. Finally, results obtained for finite gratings are compared with infinite periodic simulations and conceptual inconsistencies explained. This work  provides a consistent framework for interpreting fluorescence, absorption, and energy transfer in BIC-based periodic nanophotonic systems.
\end{abstract}

\section{Introduction}
Driven by powerful computational tools capable of modeling almost any physical system, numerical simulations have become ubiquitous in nanophotonics research.~\cite{gallinet2015methods} Unfortunately, this simplicity in producing colorful field distribution maps comes with its downside: when subtle but crucial aspects of computational electromagnetics are missed, the resulting models fail to accurately reflect the actual experimental conditions. Periodic systems are especially prone to such pitfalls: their simulation using periodic boundary conditions implies an infinite system, whilst real experiments always deal with finite structures.\cite{leveque2006optimization, pisarenko2014modal, sadrieva2017substrate, golz2024modeformation} This distinction becomes especially important when fluorescence is used in such a periodic system, as was reported recently in several experiments on bound states in the continuum (BICs).~\cite{hsu2013trapped, kodigala2017lasing, doeleman2018polarization, jin2019topologically, ho2024finiteAreaMembranes}

In the context of nanophotonics, a BIC is an eigenmode whose eigenfrequency is embedded in the free-space radiation continuum, yet which remains perfectly decoupled from it through symmetry constraints or destructive interference of the radiation channels.~\cite{marinica2008BIC,hsu2013trapped,hsu2016bicreview,azzam2018hybridbic, amrani2022friedrichWintgen} In recent years, this phenomenon has been exploited to engineer open resonant systems with strong near-field enhancement and quality factors that diverge to infinity in the ideal lossless case, while in practice a controlled perturbation converts an ideal BIC into a quasi-BIC (qBIC) --- a leaky resonance retaining an extremely high, though finite, quality factor, $Q$.~\cite{kang2023bicapplications,wang2023brillouinFolding,liang2024highQ} Common examples of such systems include photonic crystal slabs,~\cite{hsu2013trapped,zhen2014topological,jin2019topologically} metasurfaces,~\cite{koshelev2018asym,yesilkoy2019ultrasensitive} and other periodic nanostructures that support delocalized modes, i.e., modes spatially extended in at least one dimension.~\cite{bulgakov2014bloch, bulgakov2015light, yuan2017bound}
Resonant waveguide gratings are particularly attractive in this context because, being straightforward  to fabricate, they provide a technologically accessible route to delocalized guided-mode resonances and BICs over large areas and across a broad spectral range.~\cite{quaranta2018rwg,yezekyan2024qBIC} More recently, the concepts of BIC physics have been extended to hybrid plasmonic--photonic platforms, where metallic nanostructures are coupled to dielectric waveguides or metasurfaces exhibiting strong field localization in the out-of-plane direction and spectrally narrow resonances.~\cite{sadrieva2017substrate, azzam2018hybridbic,seo2020fourier, aigner2022plasmonicbic,barreda2022applications,liang2026bound, ho2026ultrathinHybridBIC} These properties make BIC-based structures particularly  promising for controlling light--matter interaction, including spontaneous emission, fluorescence, resonant absorption, exciton-photon strong coupling, and energy transfer.~\cite{seo2020fourier, zhang2023dual, kang2023bicapplications, zong2023perfectAbsorption,
qin2023strongCoupling,
wang2026anomalus} 

Indeed, strong nanophotonic resonances generally accelerate fluorescence dynamics through the Purcell effect.~\cite{girard1995Lifetime, baffou2008quenching,pelton2015modified} In the weak emitter--cavity coupling regime, the modification of the spontaneous emission rate is described by the Purcell factor $F_\textrm{P}=\frac{3}{4\pi^2}\left(\frac{\lambda_0}{n}\right)^3 \frac{Q}{V}$, where $\lambda_0$ is the free-space wavelength, $n$ the refractive index of the medium, $Q$ and $V$ are the mode quality factor and volume, respectively. However, a high-$Q$ resonance or a small mode volume do not guarantee a strong far-field fluorescence signal, which is rather determined by the balance between radiative and nonradiative decay channels, especially in lossy metallic structures.~\cite{vahala2003microcavities, kristensen2012modevolume, sauvan2013spontaneous, Stavros2025Tuning} This is particularly relevant for  structures that support BICs: in an open resonator with spatially extended qBIC modes, the conventional mode volume $V$ is not straightforwardly defined and generally requires a quasinormal-mode normalization,~\cite{kristensen2014modes, kristensen2015quasi} while the quality factor $Q$ diverges for an ideal BIC as the mode becomes decoupled from the radiation continuum.~\cite{hsu2016bicreview, koshelev2018asym}

Furthermore, in a typical fluorescence experiment, the measurable quantity --- the collected fluorescence signal --- is determined by the excitation-field enhancement at the pump wavelength combined  with the radiative emission enhancement and the collection efficiency of the optical system at the emission wavelength.~\cite{athanasiou2024alternative, fu2026air}
As a result, structures that exhibit pronounced near-field enhancement may still provide only modest radiative fluorescence enhancement when the mode volume is large, the out-coupling is inefficient, or nonradiative losses (i.e., absorption) dominate.~\cite{anger2006quenching,collison2021pslr, zhang2023dual} Indeed, it was demonstrated that fluorescence enhancement in qBIC metasurfaces can be optimized at a finite detuning from the BIC, reflecting the trade-off between the quality factor and the efficiency of radiative out-coupling.~\cite{tse2025purcellQBIC}

Even more intricate issues arise for systems that comprise two distinct fluorophores, typically referred to as \textit{donor} and \textit{acceptor}, that work in tandem.  At short donor--acceptor separation distances (typically less than $\sim$10 nm), the energy transfer between them is governed by the so-called Förster resonance energy transfer (FRET) -- a short-range, nonradiative process mediated by near-field dipole--dipole coupling.~\cite{forster1946energiewanderung, Forster1948, lakowicz2006energytransfer} At larger separation distances, the energy transfer between donor and acceptor is also possible, however it becomes photon- (or polariton-) mediated radiative transfer.~\cite{andrew2000microcavity, jones2019retreview} Nanophotonic environments can  modify both the single donor decay dynamics and the donor--acceptor interaction.~\cite{blum2012nanophotonicfret,rabouw2014photonic} This distinction is especially important in plasmonic and periodic structures, where reports of ``enhanced FRET'' may in some regimes reflect increased donor decay, improved acceptor absorption, or long-range radiative coupling, rather than the true enhancement of the near-field nonradiative Förster channel.~\cite{steele2017goldnanograting,jones2019retreview,collison2021pslr, boddeti2022longrange,yuan2024bicfret} 

The numerical modeling of these fluorescent systems is especially subtle and requires special care because a dipole in a periodic unit cell represents a coherent infinite source array rather than an isolated emitter.~\cite{Gallinet2010Accurate} Furthermore, finite-size effects, edge leakage, and the convergence of a finite grating toward the infinite-periodic limit must also be assessed explicitly before drawing conclusions about fluorescence enhancement or donor--acceptor energy transfer in finite, experimentally realistic, structures.~\cite{sadrieva2017substrate, golz2024modeformation} Recent theoretical and experimental work on nonlocal metasurfaces has shown that finite-size effects produce resonance broadening and interference fringes when the available propagation distance becomes comparable to the modal propagation length, with escape through the edges providing an additional loss channel.~\cite{hoekstra2026finitesize}

Here, we investigate fluorescence and donor--acceptor energy transfer in a hybrid plasmonic--photonic waveguide grating supporting a BIC. By comparing finite and periodic models and resolving radiative and absorptive channels, we distinguish near-field FRET from long-range transfer and acceptor-layer absorption. We show that strong BIC-related fields can enhance acceptor absorption while useful fluorescence enhancement remains moderate, and establish guidelines for interpreting dipole excitation and power normalization in reduced-dimensional simulations.

\section{Results and Discussion}
\subsection{Hybrid Plasmonic Grating Supporting a BIC}

\begin{figure}[htbp!]
\centering
\includegraphics[width=\textwidth]{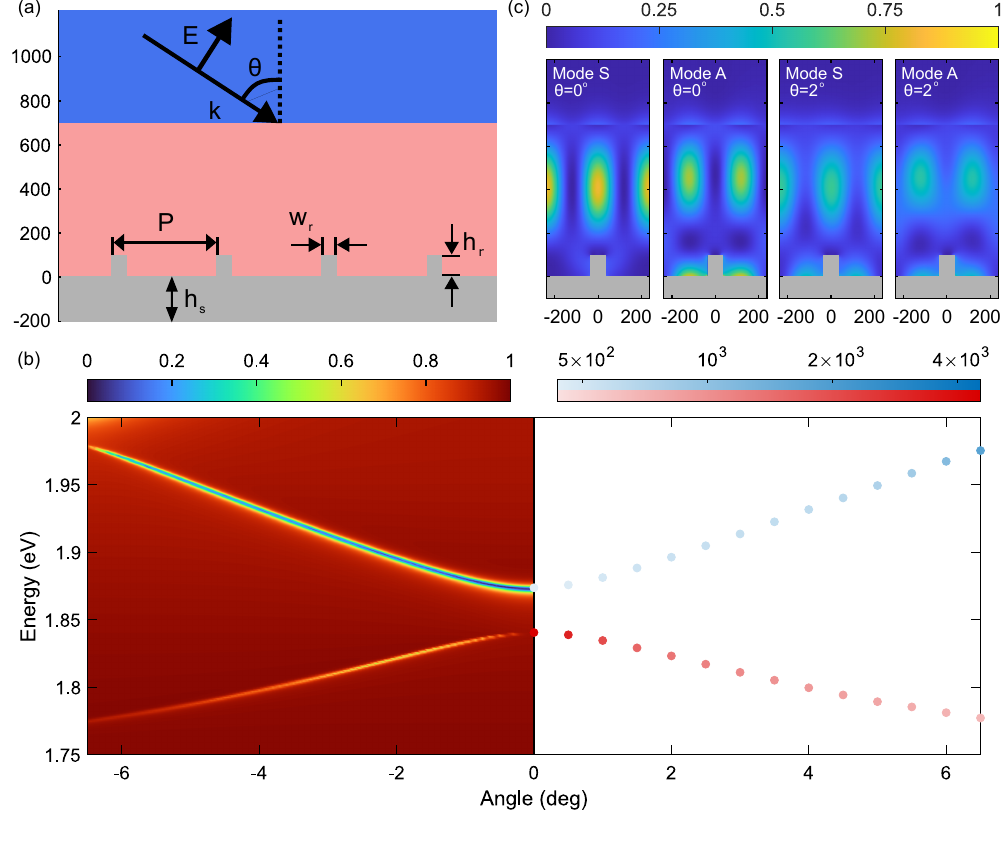}
\caption{(a) Resonant waveguide grating considered in this work: a silver grating is covered with a PVA layer and illuminated with a plane wave from the air side. The geometrical parameters are $P = 475$~nm, $w_{\rm r} = 65$~nm, $h_{\rm r} = 100$~nm, and $h_{\rm s} = 200$~nm. The thickness of the PVA waveguide layer is 700~nm. The data from Johnson and Christy are used for the permittivity of silver,~\cite{JohnsonChristy1972} while the value of $n=1.5$ is used for the refractive index of PVA. A perfect electric conductor (PEC) boundary condition was imposed at the lower surface of the silver layer without affecting the results, as the 200 nm thick layer is optically opaque at the considered wavelength range. (b) Computed angle-resolved reflection spectrum for excitation angles from $\theta = -6.5^{\circ}$ to $\theta = 0^{\circ}$. For comparison, the eigenmode-analysis results are shown for angles from $\theta = 0^{\circ}$ to $\theta = 6.5^{\circ}$. The blue and red symbols correspond to the antisymmetric mode A and the symmetric mode S, respectively. The marker transparency encodes the quality factor: higher opacity corresponds to a higher quality factor. The maximum calculated quality factor is $Q = 4561$ at the symmetry-protected BIC point near 1.84~eV. (c) Normalized electric-field modulus $|\textbf{E}|/\max(|\textbf{E}|)$ distributions obtained from eigenmode analysis for the symmetric and antisymmetric modes at $\theta = 0^{\circ}$ and $\theta = 2^{\circ}$. The horizontal axes show the position in nm.}
\label{rwg}
\end{figure}

To illustrate the possible difficulties and misinterpretations that may arise in the numerical analysis of fluorescence and energy transfer in a periodic system, we consider a resonant waveguide grating consisting of a silver (Ag) grating integrated with a PVA (polyvinyl alcohol) waveguide layer. The structure is  shown in Fig.~\ref{rwg}(a), along with the geometrical parameters and materials used in this work. Such structures support tunable BICs,~\cite{azzam2018hybridbic, seo2020fourier} as evidenced in the angle-resolved reflection spectrum shown in Fig.~\ref{rwg}(b). In addition, the mode energies obtained by an eigenmode analysis are shown on the right-hand side of Fig.~\ref{rwg}(b) for comparison. Two modes are identified in this spectral range: the antisymmetric mode (A), shown in blue, and the symmetric mode (S), shown in red. The symmetry of these modes can be understood from the distributions of the $y$-component of the electric field shown in Fig.~S1 of the Supporting Information. At normal incidence, the field distribution of mode S is symmetric about the symmetry axis of the unit cell, whereas that of mode A is antisymmetric. The opposite signs of the field-amplitude lobes visible in this panel are lost when the modulus, $|\textbf{E}|$, is calculated, Fig.~\ref{rwg}(c).

The formation of these two modes can be understood as the result of coupling between two counter-propagating waveguide modes.~\cite{Barnes1996Physical,Christ2003Waveguide, bykov2019coupled} At normal incidence, this coupling produces a bright (corresponding to A) mode and a dark (corresponding to S) mode. Although both modes exist at $\theta = 0^{\circ}$ (see Fig.~\ref{rwg}(b)), only the A mode can be excited with a plane wave, giving rise to a pronounced dip in the reflection spectrum at $\sim$1.87~eV.
In turn, the dispersive spectral feature visible at lower energy in Fig.~\ref{rwg}(b), which is associated with the excitation of mode S, vanishes at normal incidence, where the reflectance tends to unity. This corresponds to a symmetry-protected BIC at approximately 1.84~eV: a plane wave at this energy cannot in- or out-couple to the mode S. At nonzero incidence angles, the in-plane symmetry is broken, and the mode S can couple to the external radiation continuum. As a result, the reflection becomes smaller than unity, the radiative linewidth becomes finite, and the BIC transforms into a qBIC.

The normalized electric-field modulus distributions of the two modes are shown in Fig.~\ref{rwg}(c) for normal and tilted incidences. At normal incidence, mode S has a predominantly waveguide-like character, with the field mainly confined inside the PVA layer. By contrast, mode A has a hybrid waveguide--surface plasmon polariton character: a part of the field is localized inside the PVA waveguide, while another part is concentrated near the silver surface.~\cite{farhang2011plasmon} At tilted incidence, $\theta = 2^{\circ}$, the field distribution of mode A changes only marginally, whereas mode S develops an additional field contribution near the silver interface. This behavior reflects the breaking of symmetry at nonzero incidence angle, which turns the ideal BIC into a qBIC with radiative losses.

The $Q$-factor dependence of the eigenmodes is shown as a pseudocolormap in Fig.~\ref{rwg}(b). For mode S, the quality factor increases strongly as $\theta \rightarrow 0^{\circ}$, which is consistent with the suppression of radiative leakage at the symmetry-protected BIC point. However, the quality factor at $\theta = 0^{\circ}$ remains finite due to the nonradiative losses in silver. Thus, even when the radiative loss channel is suppressed, Ohmic losses in the metal still contribute to the total dissipation. In contrast, the quality factor of mode A increases gradually toward higher energies and does not exhibit the same symmetry-protected divergence at low energy when it approaches normal incidence.

The numerical simulations reported in Fig.~\ref{rwg} were performed using the \textit{Electromagnetic Waves, Wavelength Domain} interface of the Wave Optics Module in COMSOL Multiphysics\textsuperscript{\textregistered}, version 6.4 (COMSOL AB, Stockholm, Sweden).~\cite{comsol64} A two-dimensional geometry was considered, corresponding to translational invariance along the $z$--direction, perpendicular to the cross-sectional plane shown in Fig.~\ref{rwg}. The infinite grating was represented by a single unit cell of width $P$, with Bloch--Floquet periodic boundary conditions applied to its lateral boundaries.~\cite{johnson2001block, joannopoulos2008photonic} These conditions relate the electromagnetic fields in adjacent unit cells according to $\mathbf{F}(x+P,y)=\mathbf{F}(x,y)\exp(-\mathrm{i}k_xP)$, where $\mathbf{F}$ denotes either $\mathbf{E}$ or $\mathbf{H}$ and $k_x$ is the Bloch wavevector along the periodic direction. At normal incidence, $k_x=0$, and the Bloch phase factor is unity. The same boundary conditions were used to calculate the eigenmodes at the $\Gamma$ point. For tilted incidence and eigenmodes away from the $\Gamma$ point, a nonzero $k_x$ was imposed through the corresponding Floquet phase shift; for plane-wave illumination from air, $k_x=k_0\sin\theta$, where $k_0=\omega/c_0$ is the wavenumber in free space with $c_0$ being the speed of light.

\subsection{Dipole Excitation, Fluorescence, and Energy Transfer}

As discussed in the introduction, BIC-derived modes are attractive for enhancing light--matter interactions because they combine strong field localization in the waveguide layer with suppressed radiative leakage. However, an ideal symmetry-protected BIC cannot be excited by a plane wave at normal incidence. This motivates near-field excitation, for example by placing fluorescent emitters inside the waveguide layer.~\cite{abujetas2021nearfield,dyakov2021photonic,liu2023coupling,qiu2023enhancement}
Since the BIC suppresses radiative leakage, a substantial part of the energy can remain confined within the waveguide region. This suggests a possible route for enhancing energy transferred from donor to acceptor molecules embedded in the same layer. This process should be distinguished from direct FRET, in which energy is transferred through near-field donor--acceptor coupling without the emission of a propagating photon. The direct FRET rate decreases rapidly with increasing donor--acceptor separation, whereas at larger separations the acceptor can also be excited through propagation of the donor field along a BIC-derived mode.

We first consider a single dipolar emitter, modeled as an electric dipole, inside a finite resonant waveguide grating containing $\text{N}_{\rm p}=261$ periods, with the computational domain terminated by perfectly matched layers (PMLs). Details of the simulation setup are provided in the Supporting Information (Section S3.1), and the choice of $\text{N}_{\rm p}=261$ is explained in the following section. Here, we use this configuration to introduce the parameters and figures of merit relevant to our study. The effects of the boundary conditions and simulation domain size are examined in the next section.
Although the theoretical framework is applicable to other fluorophore pairs, throughout this manuscript we consider the experimentally relevant ATTO~647N--ATTO~700 donor--acceptor pair. The donor emission and acceptor absorption spectra have their strongest overlap near 1.84~eV, as shown in the Supporting Information (Section S4). The geometrical parameters were therefore chosen to place the BIC-derived resonance near this energy.

\begin{figure}[htbp]
    \centering
    \includegraphics[width=0.98\textwidth]{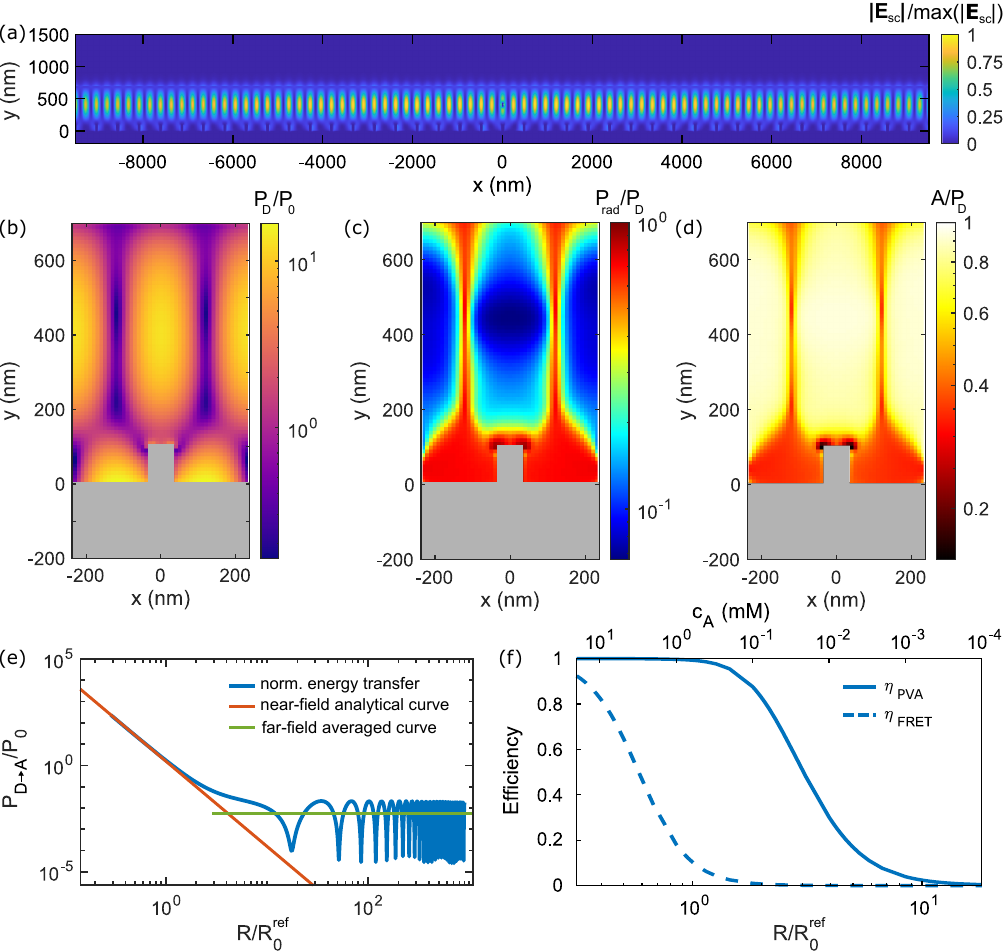}
    \caption{Finite grating with $\text{N}_{\rm p}=261$ periods under dipole excitation. (a) Normalized scattered-electric-field modulus $|\mathbf{E}_{\rm sc}|/\max(|\mathbf{E}_{\rm sc}|)$. Only the central 41 periods are shown; the normalization is $\max(|\mathbf{E}_{\rm sc}|)=1.6\times10^{5}\,\mathrm{V\,m^{-1}}$. Maps as a function of source position: (b) electromagnetic decay-rate enhancement $\text{P}_{\rm D}/\text{P}_0$; (c) radiated-power fraction $\text{P}_{\rm rad}/\text{P}_{\rm D}$; and (d) power fraction absorbed in silver, $\text{A}_{\rm silver}/\text{P}_{\rm D}$. (e) Normalized energy transfer power $\text{P}_{\rm D\rightarrow A}/\text{P}_0$ versus normalized donor--acceptor separation $R/R_{\rm 0}^{\rm ref}$ on a log--log scale. The near-field approximation, Eq.~\eqref{et},  is shown with an orange line, whereas the averaged power in the far-field is shown with a green line. (f) Acceptor-layer absorption fraction $\eta_{\rm PVA}$ versus acceptor concentration $c_{\rm A}$ (top axis) and the pair FRET efficiency evaluated at $R/R_{\rm 0}^{\rm ref}=\langle R_{\rm DA}\rangle/R_{\rm 0}^{\rm ref}$ (bottom axis), plotted using the approximate concentration--distance conversion described in the text. The horizontal axes in (e) and (f) use the distance normalization with the effective F\"orster radius $R_{\rm 0}^{\rm ref}=7.0$~nm in the uniform PVA medium.}
    \label{261per_grating}
\end{figure}

The normalized modulus of the scattered electric field produced by a single dipole is shown in Fig.~\ref{261per_grating}(a). The dipole is located at the center of the finite grating at a height of $y=410.1$~nm, corresponding to a field maximum of mode S. Its dipole moment is aligned along the $y$-axis, parallel to the dominant electric-field component of this mode. This position and orientation favor coupling to the BIC-derived resonance. The similarity between the excited field in Fig.~\ref{261per_grating}(a) and the eigenmode distribution in Fig.~\ref{rwg}(c) is consistent with excitation of mode S by the dipole.

The electromagnetic decay-rate enhancement is given by the ratio of the total power dissipated by the dipole in the structured environment to that radiated in a homogeneous reference medium:~\cite{novotny2012principles}
\begin{equation}
    \frac{\text{P}_{\rm D}}{\text{P}_0}
    =
    \frac{\operatorname{Im}\{\mathbf{p}^{*}_{\rm D}\cdot\mathbf{E}(\mathbf{r}_{\rm D})\}}
    {\operatorname{Im}\{\mathbf{p}^{*}_{\rm D}\cdot\mathbf{E}_0(\mathbf{r}_{\rm D})\}},
    \label{fl_enh}
\end{equation}
where $\mathbf{p}$ is the dipole moment, $\mathbf{E}(\mathbf{r}_{\rm D})$ is the total electric field at the source position $\mathbf{r}_{\rm D}$, and $\mathbf{E}_0(\mathbf{r}_{\rm D})$ is the reference field at the same position in an infinite homogeneous PVA medium. The source amplitude and frequency are identical in the two calculations. It is worth mentioning that $\text{P}_{\rm D}/\text{P}_0$ should not be interpreted directly as a fluorescence enhancement. The electromagnetic power balance is $\text{P}_{\rm D}=\text{P}_{\rm rad}+\text{A}_{\rm tot}$, where $\text{P}_{\rm rad}$ is the radiated power and $\text{A}_{\rm tot}$ is the absorbed power. Before acceptors are introduced, silver is the only absorbing material because the host PVA is treated as lossless.

In a homogeneous medium, the reference dipole power $\text{P}_0$ can be obtained analytically. Translational symmetry along one direction changes both the source units and the Green tensor relative to the three-dimensional case. In the present two-dimensional formulation, $\mathbf{p}_{\rm D}$ is the dipole moment per unit length, and all powers are expressed per unit length. For an in-plane dipole, $\text{P}_0=\omega^3|\mathbf{p}_{\rm D}|^2/(16c_0^2\varepsilon_0)$, as derived in the Supporting Information (Section S2.1). Combining this expression with the time-averaged source power $\text{P}_{\rm D}=(\omega/2)\operatorname{Im}\{\mathbf{p}^{*}_{\rm D}\cdot\mathbf{E}(\mathbf{r}_{\rm D})\}$ gives
\begin{equation}
    \frac{\text{P}_{\rm D}}{\text{P}_0}
    =
    \frac{8c_0^2\varepsilon_0}{\omega^2|\mathbf{p}_{\rm D}|^2}
    \operatorname{Im}\{\mathbf{p}^{*}_{\rm D}\cdot\mathbf{E}(\mathbf{r}_{\rm D})\},
    \label{fl_enh_anal}
\end{equation}
where $c_0$ is the speed of light in vacuum, $\varepsilon_0$ is the vacuum permittivity, and $\omega$ is the angular frequency.

The normalized dissipation rate at the BIC-derived resonance energy is mapped across the central grating period by scanning the dipole position in steps of 10~nm. Figure~\ref{261per_grating}(b) shows enhanced dissipation in two regions: an elliptical region inside the PVA layer and a region close to the silver surface. The elliptical region follows the field distribution of mode S and is consistent with efficient coupling to the waveguide resonance. The near-surface enhancement includes coupling to lossy fields at the metal interface and must be interpreted together with the radiative and absorptive power fractions.

The electromagnetic decay-rate enhancement in the elliptical region reaches approximately 13. Larger values occur in the immediate vicinity of the silver surface, where nonradiative dissipation can dominate. In the classical point-dipole model with a local metal response, this contribution diverges as the dipole--metal separation approaches zero. The large near-surface values therefore mainly represent metal-induced quenching and Ohmic dissipation rather than useful fluorescence enhancement.~\cite{weber1984electromagnetic,anger2006quenching} We focus on the enhancement inside the PVA layer when discussing coupling to the BIC-derived resonance, while treating the fate of the emitted energy separately.

Figure~\ref{261per_grating}(c) shows the radiated-power fraction, obtained by integrating the outward time-averaged Poynting vector over a semicircular boundary in the upper half-space above the metasurface. Its minimum is approximately 0.07, indicating that only about 7\% of the electromagnetic source power reaches this far-field boundary. This small radiative fraction is consistent with suppressed radiative leakage from the BIC-derived resonance. At positions where the dipole couples weakly to this resonance, the radiative fraction approaches unity. These positions generally correspond to weaker modal fields, as can be seen by comparing Fig.~\ref{261per_grating}(c) with Fig.~\ref{rwg}(c).

In turn, the nonradiative loss can be evaluated by integrating the fraction of power absorbed in the silver medium, which is displayed in Fig.~\ref{261per_grating}(d). As can be seen, up to $\sim$93\% of the total dipole power is absorbed in that silver layer. Interestingly, this high absorption does not occur close to the metal at the bottom of the unit cell, but rather in its middle at half-height, where the dipole energy is efficiently coupled to the BIC mode and remains largely confined inside the PVA layer whilst being unable to radiate into the far field; eventually, this energy gets absorbed in the metal. Nevertheless, as will be shown next, this effect can be exploited by introducing acceptor molecules that absorb at the donor emission wavelength, thereby opening an additional dissipation channel within the PVA layer.

In the 2D formulation, the normalized dipole--dipole energy-transfer power is
\begin{equation}
    \frac{\text{P}_{\rm D\rightarrow A}}{\text{P}_0}
    =
    \sqrt{\frac{\varepsilon_{0}}
    {\mu_{0}}}\frac{2n\sigma_{\rm A}^{\rm 2D}(\omega)}{\omega\operatorname{Im}\{\mathbf{p}^{*}_{\rm D}\cdot\mathbf{E}_{0}(\mathbf{r}_{\rm D})\}}
    \left|\mathbf{n}_{\rm A}\cdot
    \mathbf{E}_{\rm D}(\mathbf{r}_{\rm A})\right|^2
    \label{dipole}
\end{equation}
where $\mathbf{n}_{\rm A}$ is a unit vector along the acceptor dipole orientation, $n$ is the refractive index, $\sigma_{\rm A}^{\rm 2D}(\omega)$ is the absorption cross section of the acceptor, which has units of length in the 2D formulation, and $\mathbf{E}_{\rm D}(\mathbf{r}_{\rm A})$ is the donor field at the acceptor position. The difference between the 3D and 2D cases and the corresponding relation between $\sigma_{\rm A}^{\rm 3D}(\omega)$ and $\sigma_{\rm A}^{\rm 2D}(\omega)$ are discussed in the Supporting Information (Section S2.4). In case of an isolated dipole, using the derived expression for $\text{P}_{0}$ in 2D, one can get:
\begin{equation}
    \frac{\text{P}_{\rm D\rightarrow A}}{\text{P}_0}
    =
    \frac{16n\sigma_{\rm A}^{\rm 2D}(\omega)}
    {\omega^3|\mathbf{p}_{\rm D}|^2\mu_0^2c_0}
    \left|\mathbf{n}_{\rm A}\cdot
    \mathbf{E}_{\rm D}(\mathbf{r}_{\rm A})\right|^2,
    \label{dipole_sole}
\end{equation}
In a homogeneous medium, Eq.~\eqref{dipole_sole} can be approximated for the near-field regime ($nk_0R\ll 1$) as:
\begin{equation}
    \frac{\text{P}_{\rm D\rightarrow A}}{\text{P}_0}
    \approx
    \frac{4\sigma_{\rm A}^{\rm 2D}(\omega)}
    {\pi^2n^3k_0^3}\frac{1}{R^4}.
    \label{et}
\end{equation}

It is important to highlight the difference of Eq.~\eqref{et} with the original derivation by F\"orster~\cite{forster1946energiewanderung, Forster1948} that showed a $R^{-6}$ scaling of FRET for a three-dimensional molecular pair. 
In the case of two-dimensional line-dipole model, FRET exhibits a $R^{-4}$ dependence,~\cite{andrew2000microcavity} as confirmed at short separations in Fig.~\ref{261per_grating}(e), which shows the resulting energy transfer as a function of normalized donor--acceptor separation distance $R/R_0^{\rm ref}=|\mathbf{r}_{\rm D}-\mathbf{r}_{\rm A}|/R_0^{\rm ref}$, on a log--log scale. Here, $R_{0}^{\rm ref}$ is a characteristic length scale, commonly referred to as F\"orster radius, which will be discussed further. In this plot, the energy transfer was evaluated for a donor fixed in the center of the finite grating structure at $(x,y)=(0,410.1)$~nm, with its dipole moment oriented along the $y$-axis, while the acceptor was placed at the same height and scanned along the $x$-axis, with the directions of the donor and acceptor dipole moments being aligned.  At short distances, the behavior agrees with that of a homogeneous medium, the blue curve coincides with the orange one, which was calculated using Eq.~\eqref{et}. Whereas at larger distances, oscillations arise due to the periodicity of the BIC mode. These oscillations originate from the spatially extended BIC electromagnetic mode supported by the periodic structure. A similar nonlocal distance-dependent energy-transfer behaviour has been reported for systems where donor–acceptor interactions are mediated by collective photonic or lattice resonances rather than direct near-field coupling.~\cite{andrew2000microcavity,dey2026funneling}

While Eq.~\eqref{dipole} quantifies the enhancement of the energy transfer relative to the power radiated by the donor in the homogeneous medium, it does not allow assessing the energy fraction that goes into direct FRET, relative to the other competing decay channels. Hence, it is useful to introduce the FRET efficiency:
\begin{equation}
    \eta_{\rm FRET}
    =
    \frac{\text{P}_{\rm D\rightarrow A}}
    {\text{P}_{\rm D\rightarrow A}+\text{P}_{\rm D}+\text{P}_{\rm nr}},
    \label{et_efficiency}
\end{equation}
where $\text{P}_{\rm D}$ is the total electromagnetic power dissipated by the donor at its same position, orientation, and frequency in the considered environment, evaluated without the acceptor. The quantity $\text{P}_{\rm nr}$ represents the intrinsic nonradiative loss of the donor, which can be expressed using the intrinsic quantum yield $q_{\rm D}^0$ of the donor in a homogeneous reference medium:
\begin{equation}
    {\text{P}_{\rm nr}}
    =
    {\text{P}_0}\frac{1-q_{\rm D}^0}{q_{\rm D}^0}.
    \label{intrinsic_donor_loss}
\end{equation}
For ATTO~647N, we adopt the tabulated value $q_{\rm D}^0=0.65$.~\cite{FPbaseFRET} Since $\text{P}_{\rm nr}$ is not affected by the photonic environment, it contributes a constant term to the total donor power balance. 

Substituting Eqs.~\eqref{et} and \eqref{intrinsic_donor_loss} into Eq.~\eqref{et_efficiency} allows defining the aforementioned characteristic FRET length scale, $R_0$, referred to as F\"orster radius, in a way that is also consistent for a nanostructured medium. It represents the donor--acceptor separation $R = R_{0}$ at which $\eta_{\rm FRET}=50\%$, and can be expressed as:
\begin{equation}
    R_0
    =
    \left(
    \frac{\pi^2n^3k_0^3}{4\sigma_{\rm A}^{\rm 2D}(\omega)}
    \left[
    \frac{\text{P}_{\rm D}}{\text{P}_0}
    +
    \frac{1-q_{\rm D}^0}{q_{\rm D}^0}
    \right]
    \right)^{-1/4}.
    \label{forster_effective_radius}
\end{equation}
Such a definition allows us to distinguish changes in donor--acceptor transfer from changes in the competing donor decay channels. Indeed, energy transfer efficiency can increase without an increase in transfer power, for example through the inhibition of spontaneous emission.~\cite{kleppner1981inhibited,leistikow2011inhibited,rabouw2014photonic} Thus, the F\"orster radius measures an efficiency that depends both on the energy transfer to the acceptor and on the competing donor decay through radiation and absorption in the structure. Considering these quantities separately allows us to determine whether a change in FRET efficiency originates from modified donor--acceptor transfer or from changes in the competing decay channels. 

Efficient FRET requires the donor--acceptor separation $R$ to be comparable to or smaller than $R_0$. For typical molecular donor--acceptor pairs in a homogeneous medium, the F\"orster radius is below \SI{10}{nm}.~\cite{lakowicz2006energytransfer} At these separations and for molecules emitting at visible wavelengths, the near-field condition is generally satisfied, confirming the validity of the approximation. Furthermore, comparing the numerical results calculated with Eq.~\eqref{dipole} and plotted in Fig.~\ref{261per_grating}(e) with a blue curve against the analytical approximation calculated using Eq.~\eqref{et} and shown with an orange line shows a good correspondence in the small $R$ limit. This confirms that the FRET rate is not significantly modified by the nanostructured environment, as pointed out in a previous study.~\cite{blum2012nanophotonicfret} However, the FRET efficiency does get modified due to the Purcell enhancement of the donor.~\cite{collison2021pslr} 
Using Eq.~\eqref{forster_effective_radius} one can calculate the F\"orster radius $R_0^{\text{N}_{\rm p}=261}=4\pm0.1$~nm for the case of a grating with $\text{N}_{\rm p}=261$, which is significantly smaller than the  F\"orster radius in a uniform infinite PVA medium ($R_0^{\rm ref}=7$~nm), indicating lower FRET efficiency in the case of the grating at a given separation. This reduction arises from stronger competing donor decay channels, while the short-range transfer power remains nearly unchanged.

At larger separations, propagating fields and guided modes can mediate energy transfer beyond the direct near-field FRET range. In a homogeneous medium, the far-field limit $nk_{0}R\gg1$ of Eq.~\eqref{dipole_sole} gives
\begin{equation}
    \frac{\text{P}_{\rm D\rightarrow A}}{\text{P}_0}
    \simeq
    \frac{2}{\pi}\sigma_{\rm A}^{\rm 2D}(\omega)\frac{1}{R}.
    \label{et_far_field}
\end{equation}
Thus, the far-field transfer power in a homogeneous 2D medium scales as $1/R$ (Fig.~S2). In the resonant waveguide grating, the BIC-derived mode changes this distance dependence and produces oscillations associated with its spatial field distribution. The green line in Fig.~\ref{261per_grating}(e) indicates that the amplitude varies only weakly over the plotted range. This behavior reflects the spatial extent of the mode, although absorption and the finite grating size ultimately limit its propagation range.

While the previous considerations are valid for a single donor--acceptor pair, in an experiment one typically deals with a large ensemble of emitters. As we consider emitters embedded in a film, the donor--acceptor separation distance $R$ is fixed, and it is convenient to relate separation to acceptor concentration. We use the approximate geometric conversion
\begin{equation}
    c_{\rm A}\simeq
    \left(8N_{\rm A}\langle R_{\rm DA}\rangle^3\right)^{-1},
    \label{concentration_distance}
\end{equation}
where $N_{\rm A}$ is Avogadro's constant and $\langle R_{\rm DA}\rangle$ is the mean nearest-neighbour separation distance, introduced for a large ensemble of molecules. With distances in meters, this expression gives $c_{\rm A}$ in $\mathrm{mol\,m^{-3}}$, numerically equivalent to mM. It provides a characteristic concentration--distance mapping, not a full description of the nearest-neighbor distribution. We retain the F\"orster radius in PVA $R_{0}^{\rm ref}=7$~nm as the fixed plotting reference distance. Evaluating Eq.~\eqref{et_efficiency} at $R=\langle R_{\rm DA}\rangle$ gives a representative-pair estimate rather than an ensemble-averaged efficiency.

At low acceptor concentrations, direct FRET between a representative pair is weak, while an extended donor field can still be absorbed by many acceptors throughout the waveguide layer. To examine this contribution, we treat the acceptor-doped PVA as an effective absorbing medium with imaginary refractive index
\begin{equation}
    \kappa(\lambda)
    =
    \frac{\lambda}{4\pi}
    \varepsilon_{\rm A}(\lambda)c_{\rm A}\ln(10),
    \label{acceptor_extinction}
\end{equation}
where $\varepsilon_{\rm A}(\lambda)$ is the molar extinction coefficient at the calculation wavelength. ATTO~700 has a peak value of $120000\,\mathrm{cm^{-1}\,M^{-1}}$; its wavelength dependence must be retained when evaluating absorption away from the peak. Here, $\lambda$ is expressed in centimeters and $c_{\rm A}$ in M to match the units of $\varepsilon_{\rm A}$.

Compared to the previously studied situation, now in a medium with a non-vanishing imaginary part of the refractive index, the donor decay can also occur through absorption in the medium, in addition to radiation, absorption in silver or intrinsic nonradiative relaxation. The rigorous consideration of these different channels identifies where the donor energy is dissipated and is important for interpreting fluorescence and energy-transfer calculations.~\cite{seo2020fourier,collison2021pslr,yuan2024bicfret} The effective-medium calculation gives the total absorption by the acceptor layer; it does not, by itself, separate direct near-field transfer from mode-mediated absorption. Hence, it is useful to introduce the acceptor absorption efficiency:
\begin{equation}
    \eta_{\rm PVA}
    =
    \frac{\text{A}_{\rm PVA}}
    {\text{A}_{\rm PVA}+\text{A}_{\rm silver}+\text{P}_{\rm rad}+\text{P}_{\rm nr}}=
    \frac{\text{A}_{\rm PVA}}{\text{P}_{\rm tot}},
    \label{absorption_fraction}
\end{equation}
where $\text{P}_{\rm tot}$ includes radiation, absorption both in silver and in the acceptor layer, and the intrinsic nonradiative donor decay.  In other words, $\eta_{\rm PVA}$ represents the fraction of the total donor power that gets absorbed by the acceptors embedded in the PVA layer. As the acceptor concentration increases, a larger fraction of the source power gets absorbed, thereby competing with radiation and absorption in silver.

Figure~\ref{261per_grating}(f) shows $\eta_{\rm PVA}$ as a function of acceptor concentration. At $c_{\rm A}=0.02$~mM, corresponding to a characteristic nearest-neighbor distance of approximately 22~nm, the calculated $\eta_{\rm PVA}$ reaches approximately 50\%. For comparison, Eq.~\eqref{et_efficiency} gives the FRET efficiency of the donor--acceptor configuration used in Fig.~\ref{261per_grating}(e). Evaluating it at $R=\langle R_{\rm DA}\rangle$ places the pair estimate on the same approximate concentration scale. The comparison distinguishes three qualitative regimes: (i) low concentrations, where acceptor-layer absorption can be appreciable even though representative-pair FRET is weak; (ii) high concentrations, where short donor--acceptor separations favor efficient FRET; and (iii) an intermediate range, where both descriptions predict appreciable energy transfer.

\subsection{Comparison Between Infinite and Finite Gratings}

Having established the theoretical framework for quantifying radiative and nonradiative energy transfer, we examine next how the choice of numerical boundary conditions affects its calculation in periodic nanophotonic systems. For the plane-wave excitation or eigenmode analysis, an infinite periodic structure can be modeled using a single unit cell with Floquet boundary conditions consistent with the incident wavevector.~\cite{tikhodeev2002quasiguided, hsu2013trapped, quaranta2018rwg} Conversely, dipole excitation, which is used to model fluorophore emission, requires a different setting. Applying PBCs to a domain containing one dipole produces an infinite array of mutually coherent dipoles, with a relative phase between neighboring dipoles determined by the boundary conditions.~\cite{Gallinet2010Accurate} Consequently, this configuration  differs completely from an isolated donor or an ensemble of incoherently emitting molecules, as used in experiments, where a fixed phase relationship between spatially separated emitters does not exist. Rigorous treatments of individual emitters in periodic structures are nevertheless possible using Green-function formulations for localized dipole excitation, as demonstrated for bound states in the continuum in all-dielectric metasurfaces.~\cite{abujetas2021nearfield} Individual emitters and incoherent emitter ensembles can also be treated using approaches such as the diffraction matrix method, which decomposes dipole radiation into plane waves and resolves both light out-coupling and coupling to guided and plasmonic modes.~\cite{park2023fast}

\begin{figure}[htbp]
\centering
\includegraphics[width=0.98\textwidth]{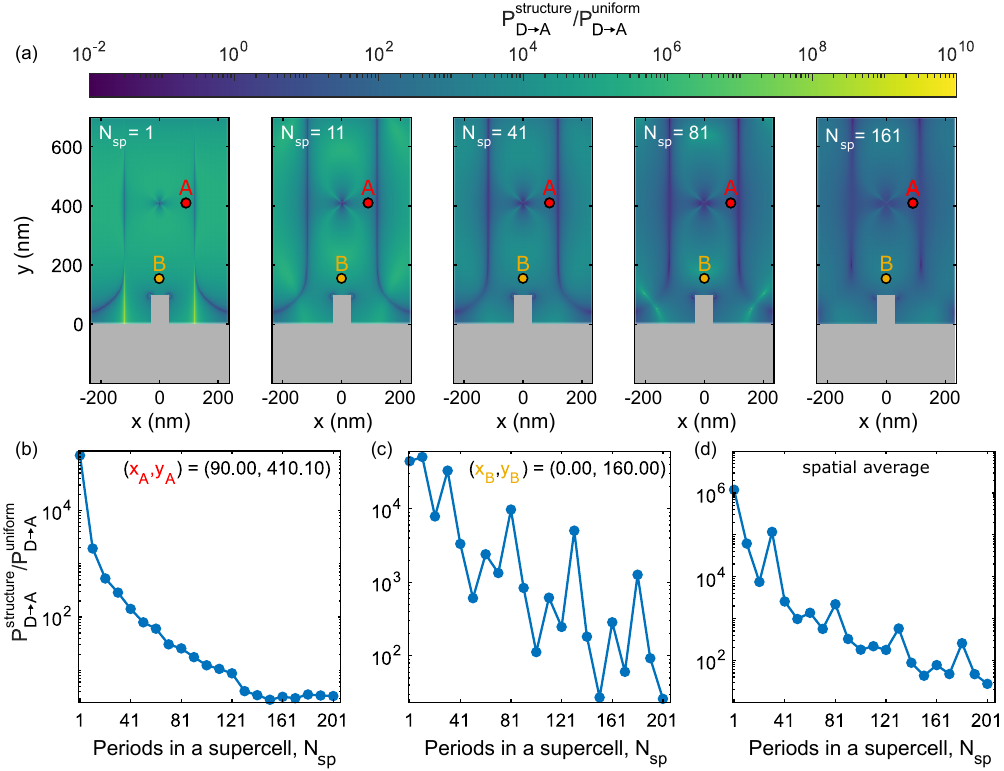}
\caption{Dependence of the calculated energy-transfer-rate enhancement on the periodic supercell size $\text{N}_{\rm sp}$. (a) Energy transfer rate spatial distributions ${\text{P}_{\rm D \rightarrow A}^{\rm structure}}/{\text{P}_{\rm D \rightarrow A}^{\rm uniform}}$ in an infinite periodic grating with $\text{N}_{\rm sp}=1$, 11, 41, 81, and 161. All maps show only the central period of the supercell and share a logarithmic color scale. The red and orange markers identify points~A and~B, respectively, which are considered in the next panels. (b,c) Enhancement as a function of $\text{N}_{\rm sp}$ at (b) point~A, $(x_A,y_A)=(90.00,410.10)\,\mathrm{nm}$, and (c) point~B, $(x_B,y_B)=(0.00,160.00)\,\mathrm{nm}$. (d) Spatial average of the local enhancement over the dielectric waveguide region within the central period. }
\label{FRET_enh_1period}
\end{figure}

Figure~\ref{FRET_enh_1period} illustrates the issues that arise in the simulation of the energy-transfer-rate enhancement,  ${\text{P}_{\rm D \rightarrow A}^{\rm structure}}/{\text{P}_{\rm D \rightarrow A}^{\rm uniform}}$, when considering dipolar sources in conjunction with PBCs. As mentioned, in a realistic experiment, one sole dipole would be placed in a periodic grating. To approach this situation, whilst still using an infinitely periodic system with PBCs, we progressively "dilute" the dipole source by considering a \textit{supercell} that includes a sole dipole in its center, but is repeated periodically using PBCs to form an infinite system. Let us first study the dependency on the number $\text{N}_{\rm sp}$ of periods included in this \textit{supercell}. 
Increasing $\text{N}_{\rm sp}$ increases the distance between periodic replicas and allows to observe variations in the interference between fields generated by the coherent dipoles, as shown in Fig.~\ref{FRET_enh_1period}(a), which plots the spatial distribution of ${\text{P}_{\rm D \rightarrow A}^{\rm structure}}/{\text{P}_{\rm D \rightarrow A}^{\rm uniform}}$ in the central period of a unit cell with $\text{N}_{\rm sp}=1, 11, 41, 81$ and $161$. Clearly, both the magnitude and spatial distribution of the enhancement vary substantially with the number of periods: for example, the pronounced features near $x\approx\pm120\,\mathrm{nm}$ for $\text{N}_{\rm sp}=1$ are strongly modified in larger supercells, while additional features appear near the grating surface for $\text{N}_{\rm sp}=81$. One may naively think that in the limit of a large supercell the enhancement would converge to a specific value, however this is not generally true and depends on the observation point. For example, enhancement at point~A, located at ($x_\textrm{A}=90, y_\textrm{A}=410\,\mathrm{nm}$), plotted in Fig.~\ref{FRET_enh_1period}(b), decreases strongly before approaching a plateau value of about 3 at $\text{N}_{\rm sp}=151$. In contrast, at point~B, ($x_\textrm{B}=0, y_\textrm{B}=160\,\mathrm{nm}$), pronounced non-monotonic oscillations accompany the overall decay (Fig.~\ref{FRET_enh_1period}(c)). The spatial average over the sampled region of the central period, shown in Fig.~\ref{FRET_enh_1period}(d), exhibits also an overall decrease with superimposed oscillations. Although the averaged enhancement shows a clear decreasing trend, the oscillations remain substantial even at large $\text{N}_{\rm sp}$. It is important to note that increasing the size of the supercell also affects the electromagnetic decay-rate enhancement $\text{P}_{\rm D}/\text{P}_0$ (i.e., Purcell effect), as discussed in Supporting Information Section S3.2. Hence, although increasing $\text{N}_{\rm sp}$ effectively changes the density of sources, such an approach cannot be reliably used to simulate varying concentrations of the emitting donor in a realistic experiment.

We next examine how the lateral extent of a finite grating affects its resonant response compared to an infinite periodic grating. Finite gratings containing $\text{N}_{\rm p}$ periods are excited by a single dipole in a domain terminated by PMLs on its sides. Note that $\text{N}_{\rm p}$ specifies the actual size of a finite grating with a single dipole source, whereas $\text{N}_{\rm sp}$ specifies the supercell size, which defines the source periodicity in an infinite grating.

\begin{figure}[htbp]
\centering
\includegraphics[width=0.98\textwidth]{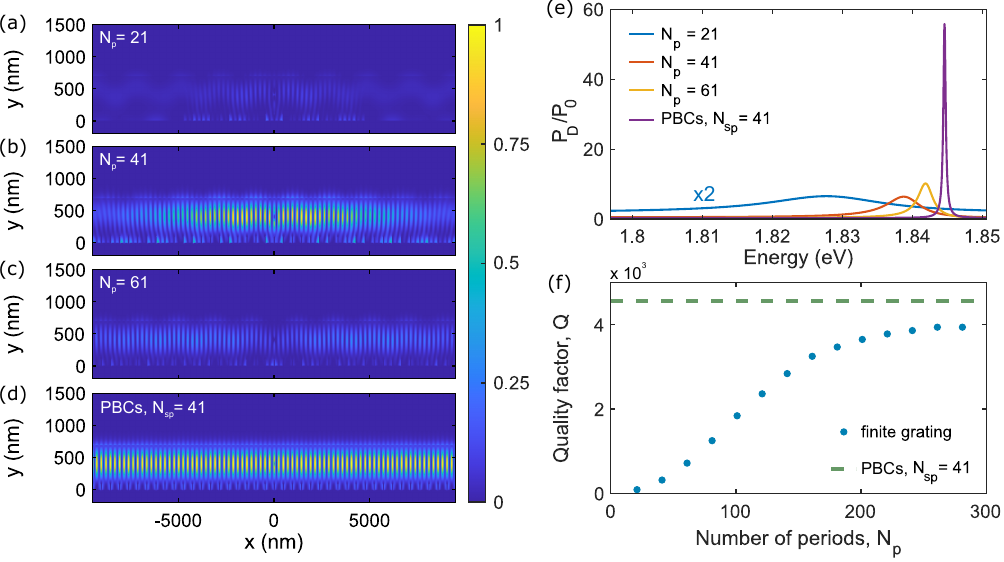}
\caption{Comparison between finite gratings under single-dipole excitation with an infinite grating under periodic dipole excitation. (a--c) The normalized scattered-electric-field modulus maps for finite gratings containing (a) $\text{N}_{\rm p}=21$, (b) 41, and (c) 61 periods, calculated using PML-terminated simulation domains. (d) Corresponding map for the infinite grating, which was modeled using PBCs with $\text{N}_{\rm sp}=41$ periods in a supercell. Every field map displays a lateral window spanning 41 structural periods; the terminations of the 61-period grating lie outside this window. The field normalization uses $\text{max}\left(|\textbf{E}_{\rm sc}|\right)=2.2\times10^5\,\mathrm{V\,m^{-1}}$ for (a--c) and $\text{max}\left(|\textbf{E}_{\rm sc}|\right)=1.79\times10^6\,\mathrm{V\,m^{-1}}$ for (d), so equal colors in the finite and periodic cases do not imply equal absolute field magnitudes. (e) Fano fit curves of the normalized power spectra, $\text{P}_{\rm D}/\text{P}_0$, for the four considered configurations. The $\text{N}_{\rm p}=21$ configuration fit is multiplied by two for legibility. (f) Quality factor extracted from the Fano fits as a function of $\text{N}_{\rm p}$ for finite gratings. The green dashed line indicates the value obtained for the infinite-periodic-grating reference.}
\label{fig:finite_infinite_comparison}
\end{figure}

Figures~\ref{fig:finite_infinite_comparison}(a--c) show the normalized scattered-electric-field modulus distributions for finite gratings with $\text{N}_{\rm p}=21$, 41, and 61. All panels use a common lateral viewing window spanning 41 structural periods. The fields exhibit the spatial modulation characteristic of the BIC-related resonance, together with a nonuniform envelope associated with the finite structure and localized excitation. The decay toward the grating terminations is particularly evident for $\text{N}_{\rm p}=21$. For $\text{N}_{\rm p}=61$, the physical terminations lie outside the displayed window. In contrast, the PBC calculation produces a field distribution that repeats from cell to cell (Fig.~\ref{fig:finite_infinite_comparison}(d)), as required by the imposed periodic excitation. The normalized scattered-field modulus is lower for $\text{N}_{\rm p}=61$ than for $\text{N}_{\rm p}=41$, showing that the local field amplitude does not increase monotonically with grating size.

Figure~\ref{fig:finite_infinite_comparison}(e) compares the normalized power spectra, $\text{P}_{\rm D}/\text{P}_0$, calculated using Eq.~\eqref{fl_enh}. A resonance associated with the BIC-derived mode is present in all four configurations. As $\text{N}_{\rm p}$ increases, the finite-grating resonance narrows and shifts to higher photon energies, toward the resonance of the infinite grating configuration. These trends concern the spectral characteristics of the resonance and do not, by themselves, establish equivalence between isolated-dipole and periodic-array emission.

Following the resonance analyses of BIC-related structures,~\cite{sadrieva2017substrate,azzam2018hybridbic} we characterize the spectra using a single-resonance Fano line shape:~\cite{Fano1961ConfigurationInteraction,Gallinet2011FanoLineShape}
\begin{equation}
    \sigma(\epsilon)=a\frac{(q+\epsilon)^2}{1+\epsilon^2}+b,
    \qquad
    \epsilon=\frac{\omega-\omega_0}{\Gamma},
    \label{fano}
\end{equation}
where $\sigma$ represents the fitted normalized power, $\omega_0$ is the resonance angular frequency, $q$ is the Fano asymmetry parameter, while $a$ and $b$ describe the amplitude and background. With this definition of $\epsilon$, $\Gamma$ is the half-width of the underlying resonant denominator, whose full width is $2\Gamma$. The corresponding quality factor is therefore $Q=\omega_0/2\Gamma$.~\cite{Fan2003FanoCMT,Christopoulos2019QualityFactor}

The extracted quality factor increases with $\text{N}_{\rm p}$ and tends toward the periodic-grating benchmark (Fig.~\ref{fig:finite_infinite_comparison}(f)). This behavior is consistent with a decreasing contribution of leakage associated with the finite lateral extent of the structure.~\cite{bulgakov2017lightEnhancement} The progressively smaller changes at large $\text{N}_{\rm p}$ indicate diminishing gains in $Q$ over the sampled range, while the finite-grating values remain below the periodic reference.~\cite{sidorenko2021accidentalBIC} Figure~\ref{fig:finite_infinite_comparison}(f) also suggests a practical guideline for selecting the grating size during device design. Increasing the number of periods beyond $\text{N}_{\rm p}=261$ yields only marginal gains in the quality factor, which may fall within the experimental uncertainty.

A word of caution is required for the normalization of power enhancement for an infinite grating, as used in Fig.~\ref{fig:finite_infinite_comparison}(e). One could be tempted to normalize the total dipole power with that of Eq.~\eqref{fl_enh_anal}. Unfortunately, this leads to wrong results since in this equation the reference power $\text{P}_{0}$ is calculated under the assumption of an isolated dipole in an infinite homogeneous space, while the reference system for this (infinitely periodic) case should be an infinitely repeated dipole. So, the reference power $\text{P}_{0}$ must be calculated in a homogeneous PVA supercell with the same periodicity, PBCs, source position, orientation, and amplitude. Matching the source configurations in this way yields an enhancement relative to the corresponding coherent array in uniform PVA. However, this normalization method does not remove the interferences between the source replicas, as it is shown in Fig.~\ref{FRET_enh_1period}.

We next analyze three representative cases in greater detail: a finite grating with $\text{N}_{\rm p}=41$ periods, corresponding to the grating size considered in the previous experimental study of this system;~\cite{seo2020fourier} a finite grating with $\text{N}_{\rm p}=261$ periods, which is already close to the infinite-grating limit; and the infinite periodic grating. For the periodic-source calculation, we use a supercell containing $\text{N}_{\rm sp}=41$ structural periods. The corresponding separation between periodically repeated dipoles is $\text{N}_{\rm sp}P=19.5~\mu\mathrm{m}$. This value was chosen as a representative diluted-source configuration with the same nominal lateral span as the experimentally motivated~\cite{seo2020fourier} $\text{N}_{\rm p}=41$ finite grating, thereby facilitating comparison between edge leakage in the finite structure and coherent source repetition in the periodic model. It is not intended to represent convergence to an isolated dipole: as shown in Fig.~\ref{FRET_enh_1period}, some local and spatially averaged quantities remain sensitive to $\text{N}_{\rm sp}$ for substantially larger supercells. Therefore, all results denoted as periodic below refer specifically to the coherent PBC source array with $\text{N}_{\rm sp}=41$, rather than to the $N_{\rm sp}\rightarrow\infty$ isolated-source limit. 

\begin{figure}[htbp]
\centering
\includegraphics[width=0.5\textwidth]{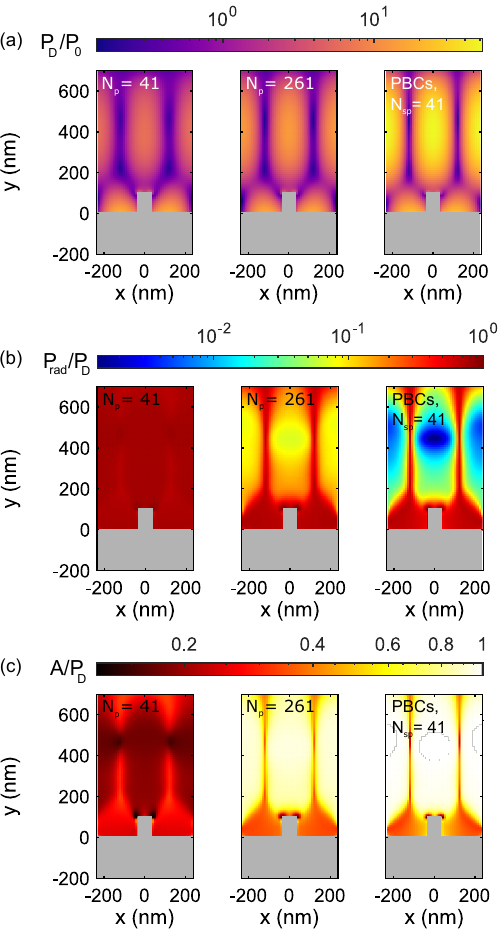}
\caption{Comparison of local power channels for finite gratings with $\text{N}_{\rm p}=41$ and $\text{N}_{\rm p}=261$ and for the infinite periodic grating with a supercell size of 41 periods. (a) Normalized total dipole power, corresponding to the electromagnetic decay-rate enhancement, for different dipole positions. (b) Fraction of the total dipole power radiated into the far field. (c) Fraction of the total dipole power absorbed in silver.}
\label{fig:finite_infinite_channels}
\end{figure}

Figure~\ref{fig:finite_infinite_channels}(a) shows the electromagnetic decay-rate enhancement calculated for varying dipole positions within the unit cell for the considered cases. In each case, the calculations were performed at the resonant energy, i.e., energy corresponding to the maximum fluorescence-rate enhancement of each configuration in Fig.~\ref{fig:finite_infinite_comparison}(e). The spatial distributions of the emission enhancement in Fig.~\ref{fig:finite_infinite_channels}(a) exhibit qualitatively similar BIC mode-like patterns in all three cases: elliptically shaped enhancement regions inside the PVA layer, away from the silver interface, and an enhancement, dominated by  absorption, in the vicinity of the metal surface. However, the magnitude of the enhancement depends strongly on the grating length. For the finite grating with $\text{N}_{\rm p}=41$, the useful enhancement inside the PVA layer is relatively weak, and the near-surface contribution is more pronounced. For $\text{N}_{\rm p}=261$, the spatial distribution becomes almost identical with that of the infinite grating, indicating that finite-size effects are substantially reduced. The maximum enhancement values in the elliptical BIC-related region inside the PVA layer are approximately 6, 13, and 52 for the gratings with $\text{N}_{\rm p}=41$, $\text{N}_{\rm p}=261$, and for the infinitely periodic grating, respectively. This confirms that the 41-period grating is still strongly affected by finite-size effects, whereas the 261-period grating is large enough to support a spatially-extended BIC mode and mimic infinitely large gratings.

Figure~\ref{fig:finite_infinite_channels}(b) shows the normalized far-field radiated power. For $\text{N}_{\rm p}=41$, the fraction of the radiated power is approximately equal to one and is nearly independent of the position in the unit cell, even at positions where the dipole couples to the BIC-like mode, as seen in Fig.~\ref{fig:finite_infinite_channels}(a). 
This is consistent with the previous observation that a significant part of the energy coupled to the guided BIC-like mode leaks out through the edges of the finite grating before being dissipated. By contrast, for $\text{N}_{\rm p}=261$ and for the infinitely periodic grating, the radiative fraction is strongly reduced in the elliptical regions of the unit cell that correspond to the maximum enhancement of emission in Fig.~\ref{fig:finite_infinite_channels}(a). Specifically, in the case of $\text{N}_{\rm p}=261$, at the position of maximum coupling to the BIC mode, only about 7\% of the total dipole power is radiated into the far field. Importantly, a small (only $\sim$0.1\%) radiative contribution remains even for the infinite periodic grating, where lateral out-coupling is absent. 

The normalized absorption in silver (i.e., the fraction of the nonradiative loss) is shown in Fig.~\ref{fig:finite_infinite_channels}(c). For $\text{N}_{\rm p}=41$, absorption in silver is reduced at positions where the dipole couples strongly to the BIC-like mode, although it remains small even outside the elliptical regions. Again, this is consistent with the fact that in this case the energy coupled to the BIC-like mode leaks to the far-field through the edges of the grating, rather than being absorbed. For $\text{N}_{\rm p}=261$ and for the infinite grating, the absorption maps are very similar: the suppression of far-field radiation leads to a larger fraction of the total power being dissipated in the metal. Overall, the analysis presented in Fig.~\ref{fig:finite_infinite_channels}  shows that the redistribution of power between radiation and absorption is controlled not only by the local modal profile but also by the finite lateral size of the grating.

\begin{figure}[htbp]
\centering
\includegraphics[width=0.98\textwidth]{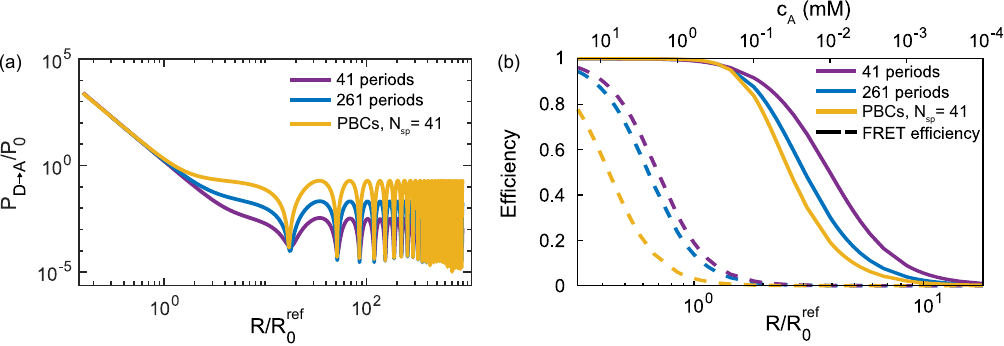}
\caption{Comparison of energy-transfer and absorption behaviors for finite gratings with $\text{N}_{\rm p}=41$ and $\text{N}_{\rm p}=261$ and for the infinite periodic grating. (a) Normalized donor--acceptor energy transfer as a function of donor--acceptor separation. (b) Absorption efficiency (solid lines) in the acceptor-doped PVA layer and direct FRET efficiency (dashed lines) as functions of acceptor concentration $c_{\rm A}$ and normalized donor--acceptor distance $R/R_{0}^{\rm ref}$ correspondingly.}
\label{fig:finite_infinite_et}
\end{figure}

Finally, let us compare the influence of a finite grating size on the energy transfer and the acceptor-layer absorption. To this end, we compare in Fig.~\ref{fig:finite_infinite_et} data for finite gratings with $N_{\rm p}=41$, $N_{\rm p}=261$ and an infinite grating with $N_{\rm sp}=41$. Panel (a) shows the normalized donor--acceptor transfer power calculated using Eq.~\eqref{dipole}. At short donor--acceptor separations, all three gratings exhibit nearly identical behaviors: the energy transfer rate is dominated by direct near-field coupling between the donor and acceptor (i.e., FRET), which occurs over nanometer-scale distances and is therefore largely insensitive to the overall grating size. 
In contrast, the efficiency and, consequently, the F\"orster radius, depends on the grating size. The calculated values are $R_{0}^{\text{N}_{\rm p}=41} = 4.9\pm 0.1$~nm, $R_{0}^{\text{N}_{\rm p}=261} = 4\pm 0.1$~nm, and $R_{0}^{\text{PBCs, }\text{N}_{\rm sp}=41} = 3.0\pm 0.1$~nm. This decrease indicates a lower FRET efficiency for a given donor-acceptor separation, since the competing donor decay channels become stronger, although the short-range energy transfer power remains nearly unchanged. At larger separations, the curves diverge: the periodic calculation exhibits the strongest oscillations, reflecting both the spatial modulation of the BIC-derived mode and interferences between the repeated dipole sources. The finite gratings exhibit a weaker long-range contribution, consistent with additional leakage from their edges. Over the plotted range, the 261-period grating follows the periodic result more closely than the 41-period grating, although this similarity does not establish an equivalence between single-dipole and coherent-array excitation!

The absorption and FRET efficiencies are compared in Fig.~\ref{fig:finite_infinite_et}(b). The direct FRET efficiency was calculated using Eq.~\eqref{et_efficiency}, the FRET efficiency curves shift toward smaller separations, consistent with the reduced effective F\"orster radii. The acceptor-layer absorption efficiency, in turn, depends on the spatial overlap of the donor field with the absorbing layer and on the competition with radiation, absorption in silver, and intrinsic donor loss. For the 41-period grating, appreciable absorption in the acceptor-doped PVA layer occurs at lower concentrations than in the 261-period or in the  periodic cases. This difference reflects changes in the field distribution and in the relative contributions of the competing decay channels; edge leakage alone does not determine the absorption efficiency. For the 261-period grating, the absorption-efficiency curve lies close to the infinite periodic result over the investigated concentration range. This agreement concerns the calculated absorption fraction and does not imply that these two models are physically equivalent.

Overall, these results demonstrate that PBCs and finite-grating simulations describe different physical situations under dipole excitation. The PBC model corresponds to an infinite coherent array of identical dipoles and is therefore not equivalent to a single emitter or to a random ensemble of emitters in a finite experimental structure. In a finite system, increasing the grating size from $\text{N}_{\rm p}=41$ to $\text{N}_{\rm p}=261$ reduces the influence of edge leakage and brings the resonance frequency and quality factor closer to the periodic-grating benchmark. However, this trend does not imply convergence of the fluorescence and energy-transfer responses, because the finite-grating calculation describes excitation by a single dipole, whereas the periodic model represents excitation by a coherent array of dipoles.

\section{Conclusions}

In this work, we analyzed the fluorescence-rate enhancement and the donor--acceptor energy transfer in a hybrid plasmonic--photonic resonant waveguide grating supporting a symmetry-protected BIC. The results show that BIC-mediated field localization can strongly modify the dipole power distribution, such that a high quality factor and large local field amplitudes do not automatically translate into useful fluorescence enhancement. For a dipole coupled to the BIC mode, radiative leakage is strongly suppressed. However, the total dissipated power is redistributed among several competing channels, including far-field radiation, Ohmic absorption in silver, and absorption in the acceptor-doped PVA layer. Therefore, the fluorescence-rate enhancement must be considered together with the underlying power balance, rather than inferred from the total decay rate alone.

Another key point is that the large decay-rate enhancement obtained in the immediate vicinity of the silver interface should not be interpreted as useful fluorescence enhancement. Although this near-surface contribution is connected with the modal field distribution, it is dominated by nonradiative coupling to the lossy metal and corresponds mainly to quenching through Ohmic dissipation. The relevant metric of the enhancement for emitters embedded in the active layer is therefore the finite BIC-related enhancement inside the PVA region, away from the metal interface. When acceptor molecules are introduced into the PVA layer, an additional dissipation channel is opened. At low acceptor concentrations, direct FRET remains weak because the mean donor--acceptor distance is larger than the F\"{o}rster radius, while absorption in the acceptor-doped PVA layer can already become significant due to the BIC field localization. At high concentrations, direct near-field FRET becomes efficient and eventually dominates the donor--acceptor interaction.

We also clarified the distinction between direct FRET and long-range BIC-mediated energy transfer. In the near-field regime, the donor-acceptor interaction follows the expected distance dependence for the 2D geometry and can be described as FRET. At larger separations, however, the interaction is mediated by a propagating (photonic or polaritonic) mode, and reabsorption through the delocalized BIC-like mode. This regime should be distinguished from direct, near-field FRET, even though it can lead to donor-acceptor energy transfer. This distinction is important for interpreting apparent FRET enhancement in resonant periodic structures, where increased acceptor excitation may originate from enhanced absorption rather than from a stronger direct F\"{o}rster channel.

Finally, we compared a finite grating with an infinite periodic model. Small finite gratings suffer from substantial radiative losses due to leakage through grating edges, which changes the balance between radiation and absorption and prevents a finite system from reproducing the infinite-grating response. As the number of periods increases, the $Q$-factor of a finite structure gradually approaches the infinite-periodic limit. However, periodic boundary conditions under dipole excitation describe an infinite coherent periodic array of identical emitters, not an isolated dipole or a random ensemble of molecules. This difference must be taken into account when using PBCs to interpret fluorescence or energy-transfer simulations. 

This work should help experimentalists working on energy-transfer enhancement and emitter--nanostructure interactions in periodic systems. It provides practical guidelines for setting up physically meaningful numerical models and avoiding pitfalls in the design of periodic systems that support BICs for fluorescence control, including donor--acceptor energy transfer.

\section*{Supporting Information}

Eigenmode analysis of the grating; analytical derivations for two-dimensional fluorescence emission and donor--acceptor energy transfer; relation between two- and three-dimensional absorption cross sections; effective-medium model for acceptor absorption; details of the finite- and periodic-grating COMSOL simulations; dependence on periodic supercell size; finite-grating cavity effects; and donor and acceptor spectra used in the calculations (PDF).
\section*{Funding}
Funding from the Swiss National Science Foundation (SNSF) under project 200021\_212758 is gratefully acknowledged.

\section*{Notes}
The authors declare no competing financial interest.


\bibliography{acs_photonics_references}

\newpage
\begin{figure}
    \centering
    {\LARGE \textbf{TOC graphic}} \par
    \vspace{10pt} 
    \includegraphics[width=8cm,height=4cm]{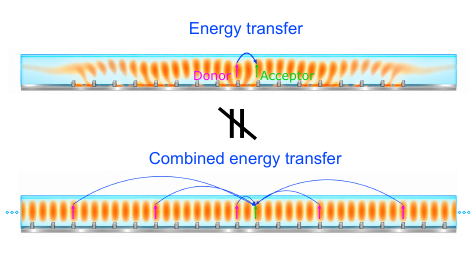}
\end{figure}

\end{document}



\setcounter{figure}{0}
\setcounter{table}{0}
\setcounter{equation}{0}

\renewcommand{\thefigure}{S\arabic{figure}}
\renewcommand{\thetable}{S\arabic{table}}
\renewcommand{\theequation}{S\arabic{equation}}

\section*{S1. Eigenmode analysis of the considered grating}

\begin{figure}[htbp!]
\centering
\includegraphics[width=\textwidth]{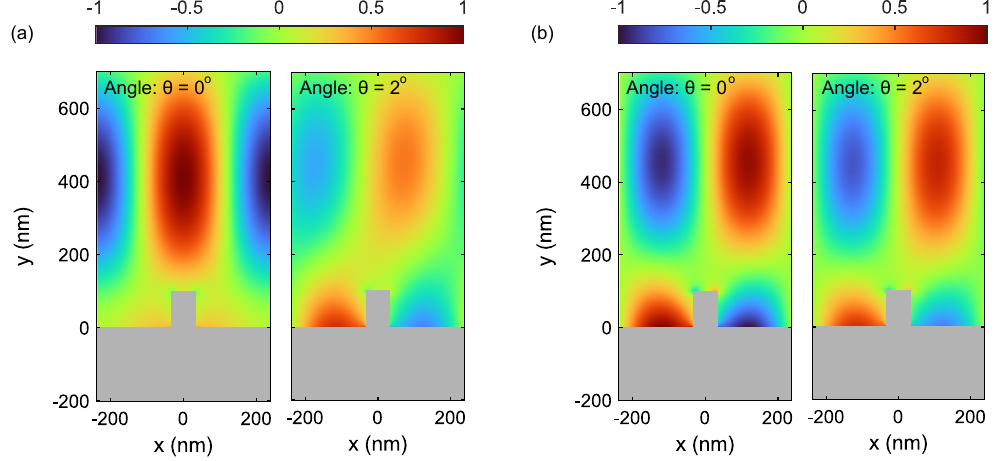}
\caption{(a) Distributions of the $y$-component of the electric field for mode S at normal and tilted incidences. The plots are normalized to the maximum value, $\max(|E_y|)=2.12\cdot 10^{8}$~V/m. (b) Corresponding field distributions for mode A, normalized to the maximum value, $\max(|E_y|)=2.1\cdot 10^{8}$~V/m.}
\label{field_y_comp_eigen}
\end{figure}

The symmetry properties of modes S and A can be seen more clearly from the $y$ component of the electric field. As shown in Fig.~\ref{field_y_comp_eigen}(a), mode S is symmetric at normal incidence, $\theta = 0^{\circ}$, and corresponds to the dark symmetry-protected BIC. By contrast, mode A is antisymmetric at normal incidence and corresponds to the bright mode (Fig.~\ref{field_y_comp_eigen}(b)). 

At tilted incidence, the in-plane symmetry of the excitation is broken. As a result, the strict symmetry protection of mode S is lifted, and the mode becomes radiatively coupled to the far field. This is reflected in the modified field distribution shown in Fig.~\ref{field_y_comp_eigen}(a). In contrast, the field distribution of mode A changes only weakly and preserves its antisymmetric character to a much larger extent (Fig.~\ref{field_y_comp_eigen}(b)). These results confirm that mode S is the symmetry-protected BIC at normal incidence and transforms into a quasi-BIC when the symmetry is broken at a nonzero incidence angle.

\section*{S2. Analytical formulas for lifetime and energy transfer calculations}

\subsection*{S2.1. Fluorescence emission power in the 2D case: analytical\\ derivation}

The fluorescence emission power of a dipole located at position $\mathbf{r}_{0}$ in 3D space is given by:~\cite{novotny2012principles}
\begin{equation}
    \text{P}_{0} = \frac{\omega^3|\mathbf{p}|^2}{2c_{0}^2\varepsilon_{0}}
    \left[\mathbf{n}_{p}\cdot\text{Im}\left\{\overleftrightarrow{\mathbf{G}}(\mathbf{r}_{0}, \mathbf{r}_{0}; \omega)\right\}\cdot\mathbf{n}_{p}\right].
\end{equation}

For a dipole oriented along the $y$-axis, only the $yy$-component of the Green's tensor contributes:
\begin{equation}
    \text{P}_{0} = \frac{\omega^3|\mathbf{p}|^2}{2c_{0}^2\varepsilon_{0}}
    \text{Im}\left\{G_{\rm yy}(\mathbf{r}_{0}, \mathbf{r}_{0}; \omega)\right\}.
    \label{poweryy}
\end{equation}

For a 2D space, assuming translation symmetry along $z$, following Ref.~\cite{martin1998}, we introduce the displacement vector:
\begin{equation}
    \mathbf{r} = \boldsymbol{\rho} - \boldsymbol{\rho}_{0}
    = [r\cos\theta,~r\sin\theta] = (x - x_{0},~y - y_{0}).
\end{equation}

If the displacement is taken along the $x$-axis, then $\theta = 0$. In this case, the $yy$-component of the 2D Green's tensor reads:~\cite{martin1998}
\begin{equation}
    G_{\rm yy}(x_{0}, x) =
    \frac{i}{4}H_{0}^{(1)}\!\left(k(x - x_{0})\right)
    - \frac{i}{4}\frac{1}{k(x - x_{0})}
    H_{1}^{(1)}\!\left(k(x - x_{0})\right),
    \label{green}
\end{equation}
where $H_{0}^{(1)}$ and $H_{1}^{(1)}$ are Hankel functions of the first kind. They can be expressed as
\begin{equation}
    H_{\nu}^{(1)}(z) = J_{\nu}(z) + iY_{\nu}(z),
    \label{hankel}
\end{equation}
with $J_{\nu}$ and $Y_{\nu}$ being Bessel functions of the first and second kinds, respectively.

Using Eqs.~\eqref{green} and \eqref{hankel}, the imaginary part of the Green's tensor component becomes
\begin{equation}
    \text{Im}\left\{G_{\rm yy}(x_{0}, x)\right\}
    = \lim_{x \rightarrow x_{0}}
    \left[
    \frac{1}{4}J_{0}\!\left(k(x - x_{0})\right)
    - \frac{1}{4}\frac{1}{k(x - x_{0})}
    J_{1}\!\left(k(x - x_{0})\right)
    \right]
    = \frac{1}{8}.
    \label{green_result}
\end{equation}

Here, the small-argument expansion of the Bessel functions ~\cite{abramowitz1948}
\begin{equation}
    J_{\alpha}(z) \sim \frac{1}{\Gamma(\alpha + 1)}\left(\frac{z}{2}\right)^{\alpha}
\end{equation}
has been used.

Substituting Eq.~\eqref{green_result} into Eq.~\eqref{poweryy}, leads to the analytical expression for the fluorescence emission power in 2D:
\begin{equation}
    \text{P}_{0} = \frac{\omega^3|\mathbf{p}|^2}{16c_{0}^2\varepsilon_{0}}.
\end{equation}
In the 2D formulation, the power has units of $\mathrm{W/m}$.

\subsection*{S2.2. Energy transfer in the 2D case: analytical derivation}

The power transferred from a donor to an acceptor dipole is given by~\cite{novotny2012principles}
\begin{equation}
    \text{P}_{\rm D \rightarrow A}
    = \frac{\omega}{2}\text{Im}\left\{\mathbf{p}_{\rm A}^{*}
    \cdot \mathbf{E}_{\rm D}(\mathbf{r}_{\rm A})\right\}.
\end{equation}

Following the derivation in Ref.~\cite{novotny2012principles}, adapted here to the 2D case, the induced dipole moment of the acceptor becomes
\begin{equation}
    \mathbf{p}_{\rm A} = \overleftrightarrow{\alpha}_{\rm A}\mathbf{E}_{\rm D}(\mathbf{r}_{\rm A}),
\end{equation}
which yields
\begin{equation}
    \text{P}_{\rm D \rightarrow A}
    = \frac{\omega}{2}\text{Im}\{\alpha_{\rm A}\}
    \left|\mathbf{n}_{\rm A}\cdot
    \mathbf{E}_{\rm D}(\mathbf{r}_{\rm A})\right|^2.
\end{equation}

The polarizability can be related to the absorption cross section, which is experimentally accessible for fluorophores:
\begin{equation}
    \sigma_{\rm A}^{\rm 2D}(\omega) = \frac{\langle \text{P}(\omega) \rangle}{\text{I}_{0}(\omega)},
\end{equation}
where $\text{I}_{0}$ is the incident intensity and $\langle \text{P}(\omega) \rangle$ is the absorbed power averaged over dipole orientations. In 2D, this averaging yields
\begin{align}
    \langle \text{P}(\omega) \rangle
    &= \frac{\omega}{2}\text{Im}\{\alpha_{\rm A}\}
    \left\langle
    \left|\mathbf{n}_{\rm A}\cdot
    \mathbf{E}_{\rm D}(\mathbf{r}_{\rm A})\right|^2
    \right\rangle \notag \\
    &= \frac{\omega}{2}\text{Im}\{\alpha_{\rm A}\}
    \frac{|\mathbf{E}_{\rm D}|^2}{2\pi}
    \int_{0}^{2\pi}\cos^2\theta~\mathrm{d}\theta \notag \\
    &= \frac{\omega}{2}\text{Im}\{\alpha_{\rm A}\}
    \frac{|\mathbf{E}_{\rm D}|^2}{2}.
\end{align}

For an acceptor dipole oriented along the $y$-axis and displaced along the $x$-axis, the energy-transfer power can be written as
\begin{equation}
    \text{P}_{\rm D \rightarrow A}
    = \sqrt{\frac{\varepsilon_{0}}{\mu_{0}}}
    n(\omega)\sigma_{\rm A}^{\rm 2D}(\omega)
    \left|E_{{\rm D},y}(x)\right|^2.
    \label{et_2D}
\end{equation}
The 2D absorption cross section $\sigma_{\rm A}^{\rm 2D}$ has units of length, so the transferred power has units of $\mathrm{W/m}$.

For a donor dipole oriented along $y$, the electric field can be written as
\begin{equation}
    E_{{\rm D},y}(x)
    = \omega^2\mu_{0}|\mathbf{p}|\,G_{\rm yy}(x_{0}, x),
    \label{efield}
\end{equation}
where $\mu_{\rm r}=1$ is assumed.

Substituting Eq.~\eqref{efield} into Eq.~\eqref{et_2D} yields
\begin{equation}
    \text{P}_{\rm D \rightarrow A}
    = \omega^4\sqrt{\varepsilon_{0}\mu_{0}^{3}}
    \,n(\omega)\sigma_{\rm A}^{\rm 2D}(\omega)
    |\mathbf{p}|^2
    |G_{\rm yy}(x_{0}, x)|^2.
\end{equation}

Using Eq.~\eqref{green} and normalizing by $\text{P}_{0}$, the normalized energy-transfer power becomes
\begin{equation}
    \frac{\text{P}_{\rm D \rightarrow A}}{\text{P}_{0}}
    = \frac{\omega}{c_{0}}n(\omega)\sigma_{\rm A}^{\rm 2D}(\omega)
    \left|
    H_{0}^{(1)}\!\left(k(x - x_{0})\right)
    - \frac{1}{k(x - x_{0})}
    H_{1}^{(1)}\!\left(k(x - x_{0})\right)
    \right|^2.
\end{equation}

For small arguments, $|x - x_{0}| \ll \lambda_{0}$, the asymptotic expansions of the Hankel functions are ~\cite{abramowitz1948}
\begin{align}
    H_{0}^{(1)}(z) &\sim \frac{2i}{\pi}\ln\left(\frac{z}{2}\right), \\
    H_{1}^{(1)}(z) &\sim -\frac{2i}{\pi z}.
\end{align}

Thus, in the near-field regime,
\begin{align}
    \frac{\text{P}_{\rm D \rightarrow A}}{\text{P}_{0}}
    &\sim \frac{\omega}{c_{0}}n(\omega)\sigma_{\rm A}^{\rm 2D}(\omega)
    \frac{4}{\pi^2}
    \left|
    \ln\left(\frac{k(x - x_{0})}{2}\right)
    + \frac{1}{[k(x - x_{0})]^2}
    \right|^2 \notag \\
    &\sim \frac{4\sigma_{\rm A}^{\rm 2D}(\omega)}{\pi^2 n^{3}(\omega) k_{0}^3}
    \frac{1}{(x - x_{0})^4}.
    \label{et_near}
\end{align}

For large arguments, $|x - x_{0}| \gg \lambda_{0}$, the Hankel function asymptotics are ~\cite{abramowitz1948}
\begin{equation}
    H_{\nu}^{(1)}(z)
    \sim \sqrt{\frac{2}{\pi z}}
    \exp\left[i\left(z - \frac{\nu\pi}{2} - \frac{\pi}{4}\right)\right].
\end{equation}

Keeping only the leading term, the far-field energy-transfer power scales as
\begin{align}
    \frac{\text{P}_{\rm D \rightarrow A}}{\text{P}_{0}}
    &\sim \frac{\omega}{c_{0}}n(\omega)\sigma_{\rm A}^{\rm 2D}(\omega)
    \left|
    \sqrt{\frac{2}{\pi k(x - x_{0})}}
    \right|^2 \notag \\
    &\sim \frac{2}{\pi}\sigma_{\rm A}^{\rm 2D}(\omega)
    \frac{1}{x - x_{0}}.
    \label{et_far}
\end{align}

\subsection*{S2.3. Energy transfer in a 2D  homogeneous medium}

To demonstrate the applicability of the derived formulas, Eqs.~\eqref{et_near} and \eqref{et_far}, a numerical simulation of a dipole embedded in a homogeneous PVA medium, $n=1.5$, was performed at a wavelength of $\lambda = 674.4$ nm. The analytical expressions, for the near-field and far-field approximations correspondingly, were then used to compare with the simulated data.

\begin{figure}[htbp]
\centering\includegraphics[width=\textwidth]{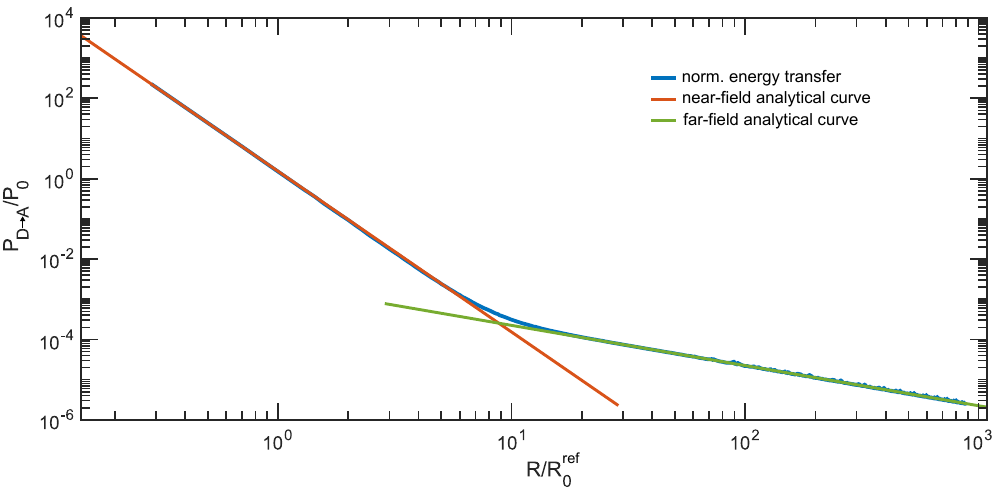}
\caption{Normalized energy transfer versus donor--acceptor distance in a homogeneous 2D medium on a log--log scale. The orange curve shows the fit using the near-field approximation, Eq.~\eqref{et_near}, while the green curve shows the fit using the far-field approximation, Eq.~\eqref{et_far}.}
\label{et_um}
\end{figure}

Figure~\ref{et_um} shows that the near-field approximation, Eq.~\eqref{et_near}, remains valid up to distances of approximately 40~nm. The far-field dependency, calculated with Eq.~\eqref{et_far}, becomes applicable from approximately 120~nm onward. Although the strict condition $|x - x_{0}| \gg \lambda_{0}$ is not yet fully satisfied at this distance, the energy-transfer dependence already begins to follow the far-field scaling.

\subsection*{S2.4. Connection between absorption cross sections in 2D and 3D}

The three-dimensional molecular absorption cross section can be
calculated from the tabulated molar extinction coefficient
$\varepsilon_{\rm A}$:~\cite{lakowicz2006energytransfer}
\begin{equation}
    \sigma_{\rm A}^{\rm 3D}(\omega)
    =
    \frac{\ln(10)}{10N_{\rm A}}\varepsilon_{\rm A}(\omega).
    \label{ac_3D}
\end{equation}
Here, $\varepsilon_{\rm A}$ is inserted in
$\mathrm{M^{-1}\,cm^{-1}}$, $N_{\rm A}$ is Avogadro's constant,
and the resulting cross section is expressed in $\mathrm{m^2}$.

We consider a homogeneous PVA reference at the vacuum wavelength
$\lambda_0=674.4$~nm. For a donor polarized along $y$ and an acceptor
with its transition dipole oriented along the same direction, the
three-dimensional energy-transfer power is~\cite{novotny2012principles}
\begin{equation}
    \text{P}_{\rm D\rightarrow A}^{\rm 3D}
    =
    \frac{3}{2}
    \sqrt{\frac{\varepsilon_0}{\mu_0}}
    n_{\rm PVA}\sigma_{\rm A}^{\rm 3D}(\omega)
    \left|E_{{\rm D},y}(x)\right|^2.
    \label{et_3D}
\end{equation}
The electric field is expressed as a peak-amplitude phasor.
The factor $1/2$ accounts for time averaging, while the factor $3$
converts the orientationally averaged molecular absorption cross
section into that of an acceptor aligned with the field.

The radiative power of the same donor in the homogeneous, lossless
reference medium is~\cite{novotny2012principles}
\begin{equation}
    \text{P}_0^{\rm 3D}
    =
    \frac{
        \omega^4|\mathbf{p}_{\rm D}|^2 n_{\rm PVA}
    }{
        12\pi c_0^3\varepsilon_0
    }.
    \label{eq:SI_P0_3D}
\end{equation}
This power excludes intrinsic nonradiative donor losses, which are
included through the reference quantum yield $q_{\rm D}^{0}$:
\begin{equation}
    \frac{\text{P}_{\rm nr}^{\rm 3D}}{\text{P}_0^{\rm 3D}}
    =
    \frac{1-q_{\rm D}^{0}}{q_{\rm D}^{0}}.
    \label{eq:SI_intrinsic_loss_3D}
\end{equation}

For the calibration, we use the orientationally averaged 3D
reference, for which
$\langle\kappa^2\rangle_{\rm or}=2/3$.
In the near-field limit, the normalized transfer power then satisfies
\begin{equation}
    \frac{
        \left\langle
            \text{P}_{\rm D\rightarrow A}^{\rm 3D}
        \right\rangle_{\rm or}
    }{
        \text{P}_0^{\rm 3D}+\text{P}_{\rm nr}^{\rm 3D}
    }
    =
    q_{\rm D}^{0}
    \frac{
        \left\langle
            \text{P}_{\rm D\rightarrow A}^{\rm 3D}
        \right\rangle_{\rm or}
    }{
        \text{P}_0^{\rm 3D}
    }
    =
    \left(\frac{R_0^{\rm 3D}}{R}\right)^6.
    \label{eq:SI_normalized_transfer_3D}
\end{equation}
The corresponding monochromatic reference F\"orster radius is
\begin{equation}
    R_0^{\rm 3D}
    =
    \left[
        \frac{
            3q_{\rm D}^{0}\sigma_{\rm A}^{\rm 3D}(\omega)
        }{
            4\pi n_{\rm PVA}^{4}k_0^{4}
        }
    \right]^{1/6},
    \qquad k_0=\frac{\omega}{c_0}.
    \label{eq:SI_Forster_radius_3D}
\end{equation}
Within this orientationally averaged pair-rate model, the transfer
efficiency equals $50\%$ at $R=R_0^{\rm 3D}$. Because the calculation
is performed at a single frequency, this radius is a monochromatic
model quantity rather than the conventional radius obtained from
the full donor--acceptor spectral
overlap.~\cite{forster1946energiewanderung}

To describe energy transfer in the two-dimensional model, we introduce
an effective absorption cross section with units of length:
\begin{equation}
    \sigma_{\rm A}^{\rm 2D}(\omega)
    =
    \frac{\sigma_{\rm A}^{\rm 3D}(\omega)}
         {L_{\rm eff}(\omega)}.
    \label{eq:SI_cross_section_conversion}
\end{equation}
The calibration length $L_{\rm eff}$ is not the physical thickness
of the acceptor layer. We determine it by requiring the fixed-orientation
2D reference pair used in our simulations to have the same
monochromatic reference F\"orster radius as the orientationally
averaged 3D model.

For the parallel in-plane dipoles considered here, the homogeneous
2D near-field expression in Eq.~\eqref{et_near} gives
\begin{equation}
    \left(R_0^{\rm 2D}\right)^4
    =
    q_{\rm D}^{0}
    \frac{
        4\sigma_{\rm A}^{\rm 2D}(\omega)
    }{
        \pi^2 n_{\rm PVA}^{3}k_0^3
    }.
    \label{eq:SI_Forster_radius_2D}
\end{equation}
Imposing $R_0^{\rm 2D}=R_0^{\rm 3D}$ therefore yields
\begin{align}
    L_{\rm eff}(\omega)
    &=
    \frac{
        4q_{\rm D}^{0}\sigma_{\rm A}^{\rm 3D}(\omega)
    }{
        \pi^2 n_{\rm PVA}^{3}k_0^3
        \left(R_0^{\rm 3D}\right)^4
    }
    \nonumber\\
    &=
    \frac{16n_{\rm PVA}k_0}{3\pi}
    \left(R_0^{\rm 3D}\right)^2.
    \label{eq:SI_Leff_pitfalls}
\end{align}
The second line follows from
Eq.~\eqref{eq:SI_Forster_radius_3D}.

For $n_{\rm PVA}=1.5$ and $R_0^{\rm 3D}\simeq7$~nm at
$\lambda_0=674.4$~nm, this relation gives
$L_{\rm eff}\simeq1.16$~nm.
The calibration matches the reference 50\%-efficiency distance;
it does not make the complete 2D and 3D transfer curves identical,
since their near-field distance dependences remain $R^{-4}$
and $R^{-6}$, respectively.
Once determined in the homogeneous reference medium, $L_{\rm eff}$
is kept unchanged in the structured-environment calculations.
Changes in the competing donor decay channels are included
separately when evaluating the transfer efficiency.

\section*{S3. Details of the numerical simulations}

\subsection*{S3.1. Setup of the COMSOL simulations}

\begin{figure}[htbp]
\centering\includegraphics[width=\textwidth]{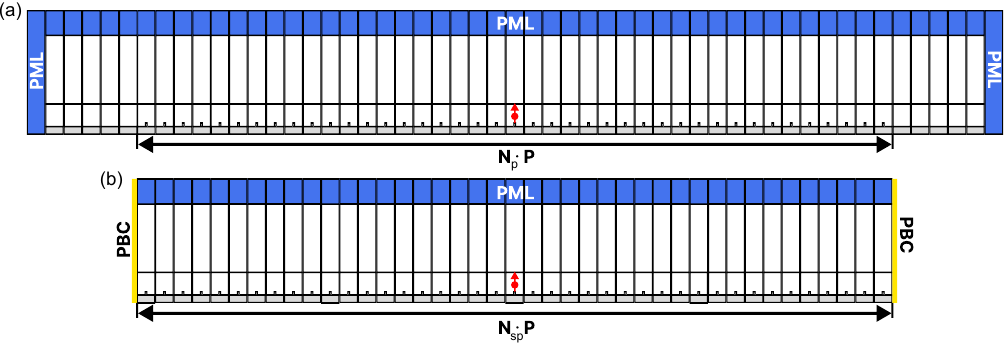}
\caption{(a) COMSOL model used for the simulation of a finite grating. (b) COMSOL model used for the simulation of the infinite grating. The infinite grating is modeled by applying periodic boundary conditions (PBCs) at the lateral boundaries.}
\label{comsol}
\end{figure}

Two different COMSOL models, illustrated in Fig.~\ref{comsol}, were used in this work. The model for a finite grating is shown in Fig.~\ref{comsol}(a). It consists of a finite structure with $\text{N}_{\rm p}$ periods and an unstructured silver surface on both sides, mimicking a realistic finite experimental sample. Perfectly matched layers (PMLs) were placed at the lateral boundaries of the simulation domain to minimize reflections.

Special care is required when defining PMLs in a layered medium consisting of silver, PVA, and air. In particular, the PMLs should be defined separately for each material layer. At the silver--PVA interface, surface plasmon polaritons (SPPs) can be excited at the grating edges and then reflected by the PMLs if they are not properly absorbed. To avoid this artificial reflection, two ports were introduced at the left and right sides of the model. A boundary-mode analysis was then added as a study step for both ports to identify the SPP mode and absorb it by mode matching.

The model for the infinite grating is shown in Fig.~\ref{comsol}(b). In this case, the simulation domain contains 41 grating periods, and periodic boundary conditions are applied at the lateral boundaries. For dipole excitation, this means that the dipole is periodically repeated every $41P$, where $P$ is the grating period. Because the considered BIC mode is spatially extended, the dipoles in this infinite periodic array can interfere with each other. This corresponds to a coherent periodic array of emitters and is generally not equivalent to the experimental situation, where the emitters are randomly positioned and randomly oriented inside the layer. A PML was used at the upper boundary of the simulation domain.

In both models, the dipole was represented by a circular domain of radius 0.1~nm polarized along the $y$-axis, as shown schematically in Fig.~\ref{comsol}. A point source was not used because it produced a singularity, making the imaginary part of the electric field at the source position strongly mesh-dependent.

\subsection*{S3.2. Fluorescence and energy-transfer rate enhancement dependence on the supercell size of the infinite periodic grating}

\begin{figure}[htbp]
\centering\includegraphics[width=\textwidth]{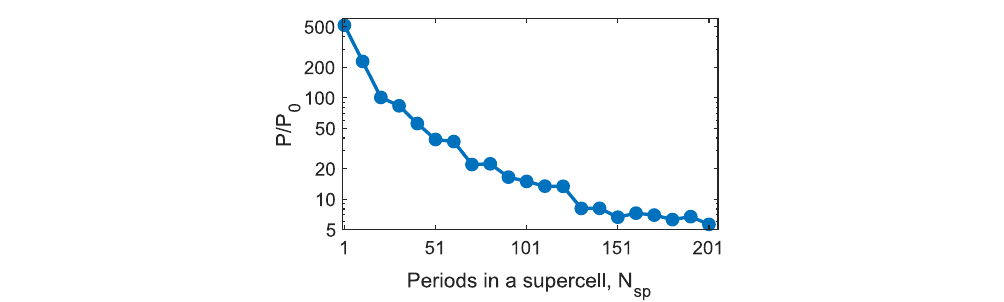}
\caption{Fluorescence-rate enhancement versus supercell size.}
\label{diff_per_fl_enh}
\end{figure}

To analyze the influence of the periodic unit-cell size, the number of periods inside a supercell $\text{N}_{\rm sp}$ in the PBC model (Fig.~\ref{comsol}(b)) was varied, and the florescence rate enhancement was calculated at the BIC wavelength, using Eq.~(1). The resulting value versus size of a supercell is demonstrated in Fig.\ref{diff_per_fl_enh}. One can see that the Purcell enhancement decreases with the number of periods in a unit cell and reaches the steady value of about 6 for high period numbers. This behavior is similar to the energy transfer enhancement illustrated in Fig.~3 of the main text.

\begin{figure}[htbp]
\centering\includegraphics[width=\textwidth]{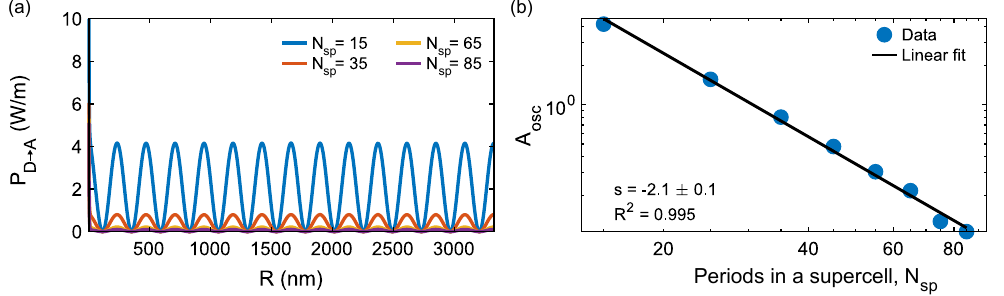}
\caption{(a) Energy transfer versus donor--acceptor distance for different numbers of periods in the PBC model. (b) Fit of the energy-transfer oscillation amplitude as a function of the number of periods in the unit cell on a log--log scale.}
\label{diff_per_osc}
\end{figure}

We also calculate the energy transfer as a function of donor--acceptor distance for different supercell sizes $\text{N}_{\rm sp}$. As shown in Fig.~\ref{diff_per_osc}(a), the amplitude of the oscillations changes with the size of the periodic unit cell. This effect arises because increasing the number of periods in the unit cell increases the separation between periodically repeated dipoles. As a result, the effective density of dipoles in the periodic array decreases as
\begin{equation}
    \rho_{\rm dip} \sim \frac{1}{\text{N}_{\rm sp}P},
\end{equation}
where $\text{N}_{\rm sp}$ is the number of structural periods in the supercell and $P$ is the grating period. Consequently, the oscillating field amplitude associated with the BIC mode excited by the repeated dipole array scales approximately as
\begin{equation}
    |E_{\rm osc}| \sim \frac{1}{\text{N}_{\rm sp}}.
\end{equation}
Since the energy-transfer power is proportional to the squared electric-field amplitude, $\text{P}_{\rm D \rightarrow A}\propto |E|^{2}$, the amplitude of the energy-transfer oscillations is expected to scale as
\begin{equation}
    A_{\rm osc} \sim \frac{1}{\text{N}_{\rm sp}^{2}}.
\end{equation}

The extracted oscillation amplitude is plotted as a function of $\text{N}_{\rm p}$ on a log--log scale in Fig.~\ref{diff_per_osc}(b). The fitted dependence is approximately $\text{N}_{\rm p}^{-2.1\pm0.1}$, in good agreement with the expected $\text{N}_{\rm p}^{-2}$ scaling for the energy-transfer power. Equivalently, if the amplitude of the electric field itself were considered instead of the energy-transfer power, an approximately $\text{N}_{\rm p}^{-1}$ scaling would be expected.

\subsection*{S3.3. Cavity effect for finite gratings}

A finite grating, illustrated in Fig.~\ref{comsol}(a), acts as a lateral cavity. Since the silver surface is unstructured outside the grating, the mode excited by the dipole inside the grating can be partially reflected from the grating edges. To demonstrate this effect, simulations were performed for finite gratings with different numbers of periods.

\begin{figure}[htbp]
\centering\includegraphics[width=\textwidth]{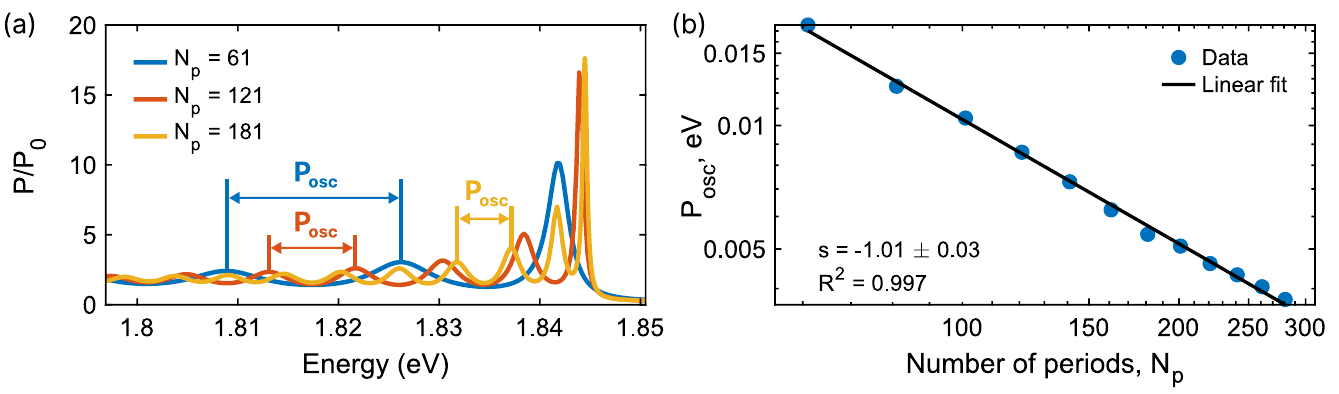}
\caption{(a) Fluorescence enhancement versus energy for finite gratings with different period numbers. The oscillations caused by the lateral cavity effect are clearly visible. The spectral periods of these oscillations, $\text{P}_{\rm osc}$, are marked in the plot with the corresponding colors. (b) Fit of the oscillation period as a function of the grating size, $\text{N}_{\rm p}P$, on a log--log scale. The fitted dependence agrees with the expected scaling, $\text{P}_{\rm osc}\sim 1/(\text{N}_{\rm p}P)$.}
\label{cavity}
\end{figure}

Figure~\ref{cavity}(a) shows the fluorescence enhancement as a function of energy for finite gratings with different sizes. In addition to the main resonance peak associated with the excited BIC mode, smaller oscillatory peaks are observed. These oscillations have a characteristic spectral period, $\text{P}_{\rm osc}$, which decreases as the grating size increases.

If these oscillations originate from a lateral cavity effect, their spectral period should be inversely proportional to the lateral size of the grating:
\begin{equation}
    \text{P}_{\rm osc}\sim\frac{1}{\text{N}_{\rm p}P}.
\end{equation}
The oscillation period is plotted on a log--log scale as a function of the grating size in Fig.~\ref{cavity}(b). The fitted slope is close to $-1$, confirming the expected inverse scaling and supporting the interpretation of the smaller peaks as finite-size cavity oscillations.

\section*{S4. Considered donor--acceptor pair}

\begin{figure}[htbp!]
\centering\includegraphics[width=0.5\textwidth]{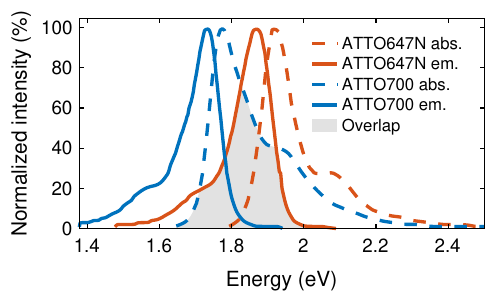}
\caption{Absorption and emission spectra of ATTO 647N and ATTO 700 used for the spectral-overlap analysis. The spectra were obtained from the FPbase FRET Calculator and interpolated onto a common photon-energy grid for use in the calculations.~\cite{Lambert2019FPbase,FPbaseFRET}}
\label{em_abs_spectr}
\end{figure}

In our work we used the ATTO 647N--ATTO 700 pair. The emission and absorption spectra of both emitters are shown in Fig.~\ref{em_abs_spectr}. The maximum molar extinction coefficient of ATTO 700 is $\varepsilon = 120000\ \text{cm}^{-1}\text{M}^{-1}$ and it was used for all the simulations. The spectral data used for ATTO 647N and ATTO 700 were taken from the FPbase FRET Calculator. ~\cite{Lambert2019FPbase,FPbaseFRET} The spectra were converted from wavelength to photon energy using $E = hc/\lambda$ and linearly interpolated onto a common energy grid for the overlap analysis.

\bibliography{acs_photonics_references}